\documentclass[final,5p,times,twocolumn]{elsarticle}

\graphicspath{{\subfix{Figures/}}}
\usepackage{lineno}
\modulolinenumbers[1]
\usepackage{graphicx}
\usepackage{siunitx}
\usepackage{upgreek}
\usepackage{subfig}
\usepackage{comment}
\usepackage{multirow}
\usepackage{tabularx}
\usepackage[version=4]{mhchem}
\usepackage{amsmath}

\usepackage[svgnames]{xcolor}
\usepackage[colorlinks,linktocpage]{hyperref}
\AtBeginDocument{%
    \hypersetup{
        citecolor=Red,
        linkcolor=Green,
        urlcolor=Blue}}

\journal{NIM-A}

\usepackage{subfiles} 

\begin{document}

\begin{frontmatter}

\title{Measured and projected beam backgrounds \\in the Belle~II experiment at the SuperKEKB collider}

\author[instHawaii]{A.~Natochii\corref{cor1}}\ead{natochii@hawaii.edu}%
\author[instHawaii]{T.~E.~Browder}%
\author[instDESY]{L.~Cao}%
\author[instTriesteINFN,instTriesteElettra]{G.~Cautero}%
\author[instDESY]{S.~Dreyer}%
\author[instPhysGeorgAugustUniv]{A.~Frey}
\author[instTriesteUNIV,instTriesteINFN]{A.~Gabrielli}%
\author[instTriesteINFN,instTriesteElettra]{D.~Giuressi}%
\author[instKEK]{T.~Ishibashi}%
\author[instTriesteINFN]{Y.~Jin}%
\author[instNagoya]{K.~Kojima}%
\author[instMPP]{T.~Kraetzschmar}%
\author[instTriesteINFN]{L.~Lanceri}%
\author[instHiroshima]{Z.~Liptak}%
\author[instWayneState,instKEK]{D.~Liventsev}%
\author[instIFIC]{C.~Marinas}%
\author[instPisaUNIV,instPisaINFN]{L.~Massaccesi}%
\author[instKEK,instSOKENDAI,instNagoya]{K.~Matsuoka}%
\author[instDuke]{F.~Meier}%
\author[instVictoria]{C.~Miller}%
\author[instKEK,instSOKENDAI]{H.~Nakayama}\ead{hiroyuki.nakayama@kek.jp}%
\author[instDESY]{C.~Niebuhr}%
\author[instLjubljanaJSI]{A.~Novosel}%
\author[instDuke]{K.~Parham}
\author[instMPP]{I.~Popov}%
\author[instPisaUNIV,instPisaINFN]{G.~Rizzo}%
\author[instVictoria]{J.~M.~Roney}%
\author[instRCNP]{S.~Y.~Ryu}%
\author[instLjubljanaUniLJ,instLjubljanaJSI]{L.~Santelj}%
\author[instDuke]{S.~Schneider}
\author[instHawaii]{J.~Schueler}%
\author[instPhysGeorgAugustUniv]{B.~Schwenker}%
\author[instKEK]{X.~D.~Shi}%
\author[instMPP]{F.~Simon}%
\author[instDESY]{S.~Stefkova}%
\author[instDESY]{M.~Takahashi}%
\author[instUTokyo]{H.~Tanigawa}%
\author[instKEK]{N.~Taniguchi}%
\author[instKEK]{S.~Terui}%
\author[instHawaii]{S.~E.~Vahsen}\ead{sevahsen@hawaii.edu}%
\author[instTriesteUNIV,instTriesteINFN]{L.~Vitale}%
\author[instDuke]{A.~Vossen}
\author[instUTokyo]{Z.~Wang}%
\author[instKrakow]{J.~Wiechczynski}%
\author[instMPP]{H.~Windel\corref{aff2}}%
\author[instNagoya]{K.~Yoshihara}%


\address[instHawaii]{University of Hawaii, Honolulu, Hawaii 96822, USA}
\address[instDESY]{Deutsches Elektronen--Synchrotron, 22607 Hamburg, Germany}
\address[instTriesteINFN]{INFN Sezione di Trieste, I-34127 Trieste, Italy}
\address[instTriesteElettra]{Elettra Sincrotrone Trieste SCpA, I-34149 Trieste, Italy}
\address[instPhysGeorgAugustUniv]{II. Physikalisches Institut, Georg-August-Universit\"at G\"ottingen, 37073 G\"ottingen, Germany}
\address[instTriesteUNIV]{Dipartimento di Fisica, Universit\`{a} di Trieste, I-34127 Trieste, Italy}
\address[instKEK]{High Energy Accelerator Research Organization (KEK), Tsukuba 305-0801, Japan}
\address[instNagoya]{Graduate School of Science, Nagoya University, Nagoya 464-8602, Japan}
\address[instMPP]{Max-Planck-Institut f\"{u}r Physik, 80805 M\"{u}nchen, Germany}
\address[instHiroshima]{Hiroshima University, Higashi-Hiroshima, Hiroshima 739-8530, Japan}
\address[instWayneState]{Wayne State University, Detroit, Michigan 48202, U.S.A.}
\address[instIFIC]{Instituto de Fisica Corpuscular, Paterna 46980, Spain}
\address[instPisaUNIV]{Dipartimento di Fisica, Universit\`{a} di Pisa, I-56127 Pisa, Italy}
\address[instPisaINFN]{INFN Sezione di Pisa, I-56127 Pisa, Italy}
\address[instSOKENDAI]{The Graduate University for Advanced Studies (SOKENDAI), Hayama 240-0193, Japan}
\address[instDuke]{Duke University, Durham, North Carolina 27708, U.S.A.}
\address[instVictoria]{University of Victoria, Victoria, British Columbia, V8W 3P6, Canada}
\address[instLjubljanaJSI]{J. Stefan Institute, 1000 Ljubljana, Slovenia}
\address[instRCNP]{Research Center for Nuclear Physics, Osaka University, Osaka 567-0047, Japan}
\address[instLjubljanaUniLJ]{Faculty of Mathematics and Physics, University of Ljubljana, 1000 Ljubljana, Slovenia}
\address[instUTokyo]{Department of Physics, University of Tokyo, Tokyo 113-0033, Japan}
\address[instKrakow]{H. Niewodniczanski Institute of Nuclear Physics, Krakow 31-342, Poland}

\cortext[cor1]{Corresponding author}
\cortext[aff2]{Now at Clinic for Cardiothoracic and Vascular Surgery, University Medical Center, 37075 G\"{o}ttingen, Germany}

\begin{abstract}
The Belle~II experiment at the SuperKEKB electron-positron collider aims to collect an unprecedented data set of $50~{\rm ab}^{-1}$ to study $CP$-violation in the $B$-meson system and to search for Physics beyond the Standard Model. SuperKEKB is already the world's highest-luminosity collider. In order to collect the planned data set within approximately one decade, the target is to reach a peak luminosity of \SI{6e35}{cm^{-2}.s^{-1}} by further increasing the beam currents and reducing the beam size at the interaction point by squeezing the betatron function down to $\beta^{*}_{\rm y}=\SI{0.3}{mm}$. To ensure detector longevity and maintain good reconstruction performance, beam backgrounds must remain well controlled. We report on current background rates in Belle~II and compare these against simulation. We find that a number of recent refinements have significantly improved the background simulation accuracy. Finally, we estimate the safety margins going forward. We predict that backgrounds should remain high but acceptable until a luminosity of at least \SI{2.8e35}{cm^{-2}.s^{-1}} is reached for $\beta^{*}_{\rm y}=\SI{0.6}{mm}$. At this point, the most vulnerable Belle~II detectors, the Time-of-Propagation (TOP) particle identification system and the Central Drift Chamber (CDC), have predicted background hit rates from single-beam and luminosity backgrounds that add up to approximately half of the maximum acceptable rates. 

\end{abstract}


\end{frontmatter}



\section{\label{sec:Introduction}Introduction}

The Belle~II experiment~\cite{Belle2TDR2010,ADACHI201846} studies $CP$-violation in the $B$-meson system and searches for Physics beyond the Standard Model, including evidence of dark sector particles,  in decays of $B$-mesons, $D$-mesons and tau leptons~\cite{Kou2019}. The SuperKEKB collider~\cite{Ohnishi2013} produces particles of interest by colliding electron and positron beams with asymmetric energies, mainly at the $\Upsilon(4S)$ resonance. SuperKEKB is a major upgrade of KEKB~\cite{nlhep1995,Kurokawa2003,Abe2013} and has been operational since 2016. The machine has already reached a world-record luminosity of \SI{4.65e34}{cm^{-2}.s^{-1}} with a vertical betatron function of $\beta^{*}_{\rm y}=\SI{1.0}{mm}$ at the interaction point (IP), but the goal is to increase the luminosity by another order of magnitude in the coming decade, with a current target peak luminosity of \SI{6e35}{cm^{-2}.s^{-1}} for $\beta^{*}_{\rm y}=\SI{0.3}{mm}$. Luminosity increases by increasing beam currents and reducing the beam size at the IP, utilizing low-emittance colliding beams and the so-called nano-beam scheme~\cite{Baszczyk2013}.

Beam particles that deviate from the nominal orbit are eventually lost by hitting the beam pipe's inner wall or other machine apparatus. If the loss position is close to the Belle~II, generated shower particles might reach the detector and increase its dose rate and hit rate. This increase is referred to as ``beam (induced) background'' and is one of the most difficult challenges at SuperKEKB. In the SuperKEKB and Belle~II designs, it was estimated that several Belle~II sub-detectors would be subject to close-to-tolerable backgrounds at the target peak luminosity~\cite{Nakayama2015,LEWIS201969}. The most vulnerable sub-detectors are the Time-of-Propagation (TOP) particle identification system and the Central Drift Chamber (CDC). 
In the TOP, higher hit rates increase the accumulated output charge in the micro-channel-plate photo-multiplier tubes (MCP-PMTs) used to read out Cherenkov photons propagated in quartz bars, which can degrade the quantum efficiency of the PMTs.
One key issue in the CDC is that pattern recognition of charged tracks becomes increasingly difficult as the wire-hit rate increases.



Given the importance of beam background mitigation to the experiment's success, we have studied such backgrounds extensively in the early stages of SuperKEKB running. The Belle~II/SuperKEKB project has three major commissioning phases: 

\begin{itemize}
    
    \item {\it Phase~1} was carried out in Spring 2016. No beam collisions occurred, as SuperKEKB was running without the final focusing system. Belle~II had not yet been installed at the IP. Instead, a system of dedicated beam background detectors, collectively known as BEAST~II, was placed around the IP. We found that the background level around the IP was safe for Belle~II to be installed. Results of the Phase~1 measurements and simulation are reported in Ref.~\cite{LEWIS201969}.
    
    \item {\it Phase~2} began in March 2018 and concluded in July 2018. The machine group demonstrated first the $e^{+}e^{-}$ collisions with Belle~II (except for the vertex detector) now installed at the IP. This commissioning phase confirmed that proceeding and installing the sensitive vertex detector was safe. Details and results of the Phase~2 beam background study can be found in Ref.~\cite{Liptak2021}. 
    
    \item {\it Phase~3}, which started in March 2019, is dedicated to physics data taking with a fully instrumented Belle~II detector and to increasing the instantaneous luminosity above \SI{1e35}{cm^{-2}.s^{-1}}. We aim to accumulate \SI{50}{ab^{-1}} of data by the 2030s, anticipating 7--8~months of operation per year, and assuming 70\% of that operation time is spent on physics runs~\cite{LumiProjSuperKEKB}.
\end{itemize}


In the rest of Section~\ref{sec:Introduction}, we describe the SuperKEKB collider and the Belle~II detector and provide an overview of the main beam-induced background sources, background countermeasures, and relevant beam instrumentation. Section~\ref{sec:CurrentBackgroundLevelsAndMargin} reports on the current (early Phase~3) background levels and safety margins of the Belle~II sub-detectors. In Section~\ref{sec:BackgroundSimulation}, we describe the beam background Monte-Carlo (MC) simulation methodology. In Section~\ref{sec:MeasuredBackgroundComposition}, we explain the methodology of background measurements and modeling. Section~\ref{sec:SummaryOfCurrentBackgroundLevels} reports on the measured background composition in Belle~II. In Section~\ref{sec:SummaryOfCurrentBackgroundLevels} we also apply correction factors of each simulated background process in each sub-detector to enforce a full agreement with measurements. This detailed model is required to reliably extrapolate the current backgrounds to different beam conditions. Section~\ref{sec:Extrapolations} describes an extrapolation of backgrounds towards higher luminosity and provides expected detector safety factors\footnote{The safety factor is defined as a ratio between the detector limit and predicted background rate. It shows how much the background level can increase before reaching the detector limit.}. Finally, in Section~\ref{sec:Conclusion}, we summarize and discuss our findings and their implications.

\subsection{SuperKEKB and Belle~II}
\label{subsec:Belle2Experiment}

Here, we briefly review the collider and detector sub-systems involved in the beam-induced background analysis. Further details can be found in Refs.~\cite{Belle2TDR2010,ADACHI201846,Ohnishi2013}.

\begin{figure}[htbp]
\centering
\includegraphics[width=\linewidth]{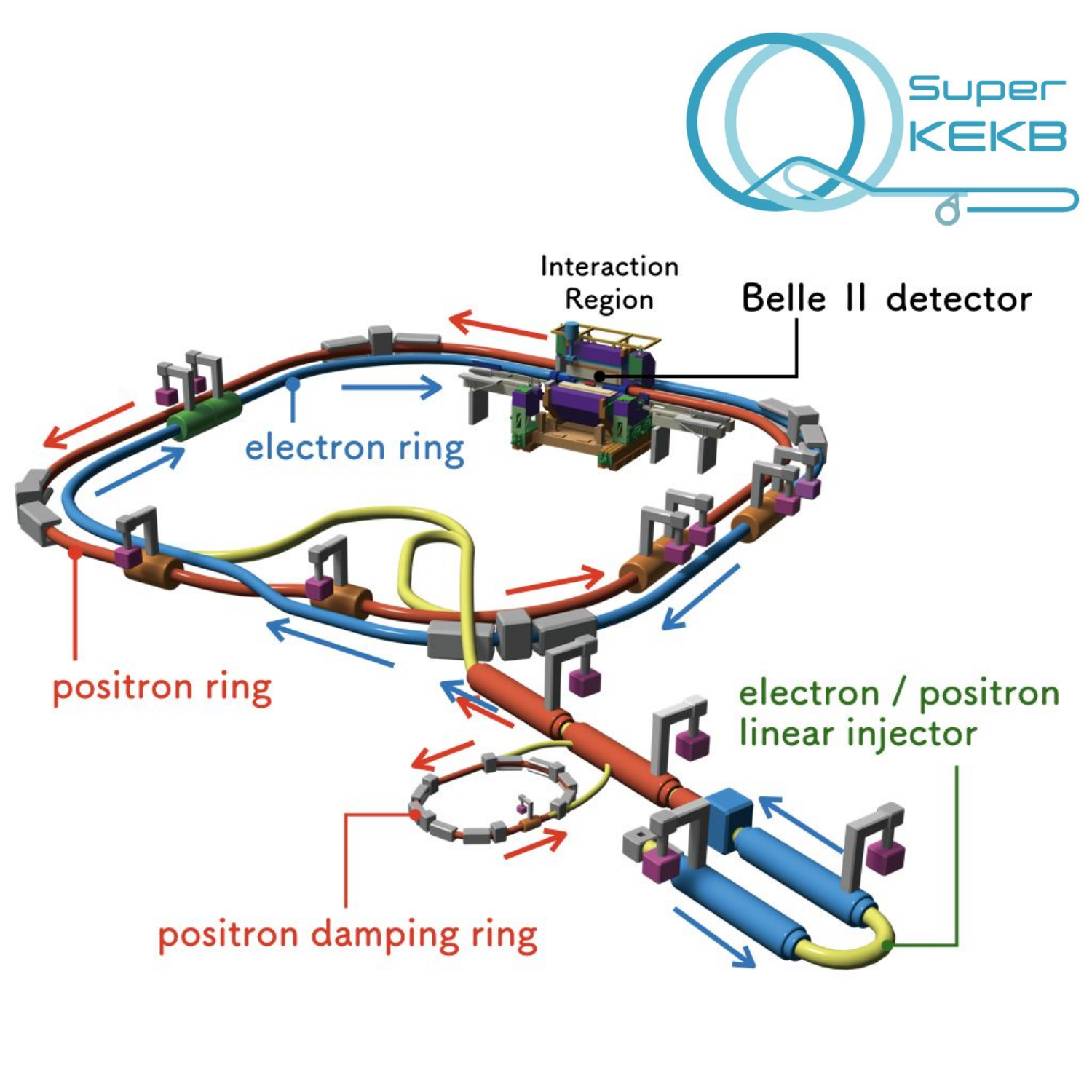}
\caption{\label{secIntro:fig3}Schematic drawing of the SuperKEKB collider.}

\end{figure}

\noindent {\it SuperKEKB}, illustrated in Fig.~\ref{secIntro:fig3}, is an upgrade of the KEKB accelerator. It is a \SI{3}{km}-circumference asymmetric-energy electron-positron collider with a center-of-mass (CM) energy of $\sqrt{s} = \SI{10.58}{GeV}$ which corresponds to the mass of the $\Upsilon(4S)$ resonance. At the IP, \SI{7}{GeV} electrons stored in the high-energy ring (HER) collide with \SI{4}{GeV} positrons accumulated in the low-energy ring (LER). To reach collision luminosity of the order of \SI{1e35}{cm^{-2}.s^{-1}}, SuperKEKB utilizes the so-called nano-beam scheme, where the vertical and horizontal beam sizes at the IP are squeezed down to \SI{\sim 50}{nm} and \SI{\sim 10}{\micro m}, respectively, with a horizontal crossing angle of \SI{83}{mrad} to avoid the hour-glass effect. The relatively large crossing angle also allows i) a new final focusing system with superconducting quadrupole magnets (QCS) to reside closer to the IP, ii) separate beamlines for the HER and LER, and iii) a design that avoids combined-function IP magnets. 
To eliminate luminosity degradation caused by beam-beam resonances, dedicated sextupole magnets are used for the Crab-Waist collision scheme implementation~\cite{raimondi2007beambeam}, which aligns the vertical waistline of one beam along the trajectory of the other beam at the IP.

The upgrade from KEKB to SuperKEKB included the following major items. We note that the list is not exhaustive.
\begin{itemize}
    \item Short LER dipole magnets were replaced with longer ones.
    \item The interaction region (IR), $\pm\SI{4}{m}$ around the IP, was redesigned. This region hosts the Belle~II detector, the final focusing system and the IR beam pipe assembly.
    \item Beam pipes with a titanium nitride (TiN) coating and antechambers were installed in the LER to reduce the power density of the synchrotron radiation (SR) and to suppress electron-cloud formation.
    \item A damping ring (DR) was constructed to reduce the injected positron beam emittance.
    \item The radio-frequency (RF) system was modified to enable higher beam currents.
    \item The collimation system was upgraded, see Section~\ref{subsec:BackgroundMitigation}.
\end{itemize}

There are two major upgrades of SuperKEKB planned in the next decade, during Long Shutdown~1 (LS1), which began in July~2022, and during Long Shutdown~2 (LS2), expected to begin around 2027. Possible future upgrades of the detector are strongly linked to upgrades of the machine. The most crucial upgrades under consideration are discussed in Ref.~\cite{Natochii2022}.

\begin{figure}[htbp]
\centering
\includegraphics[width=\linewidth]{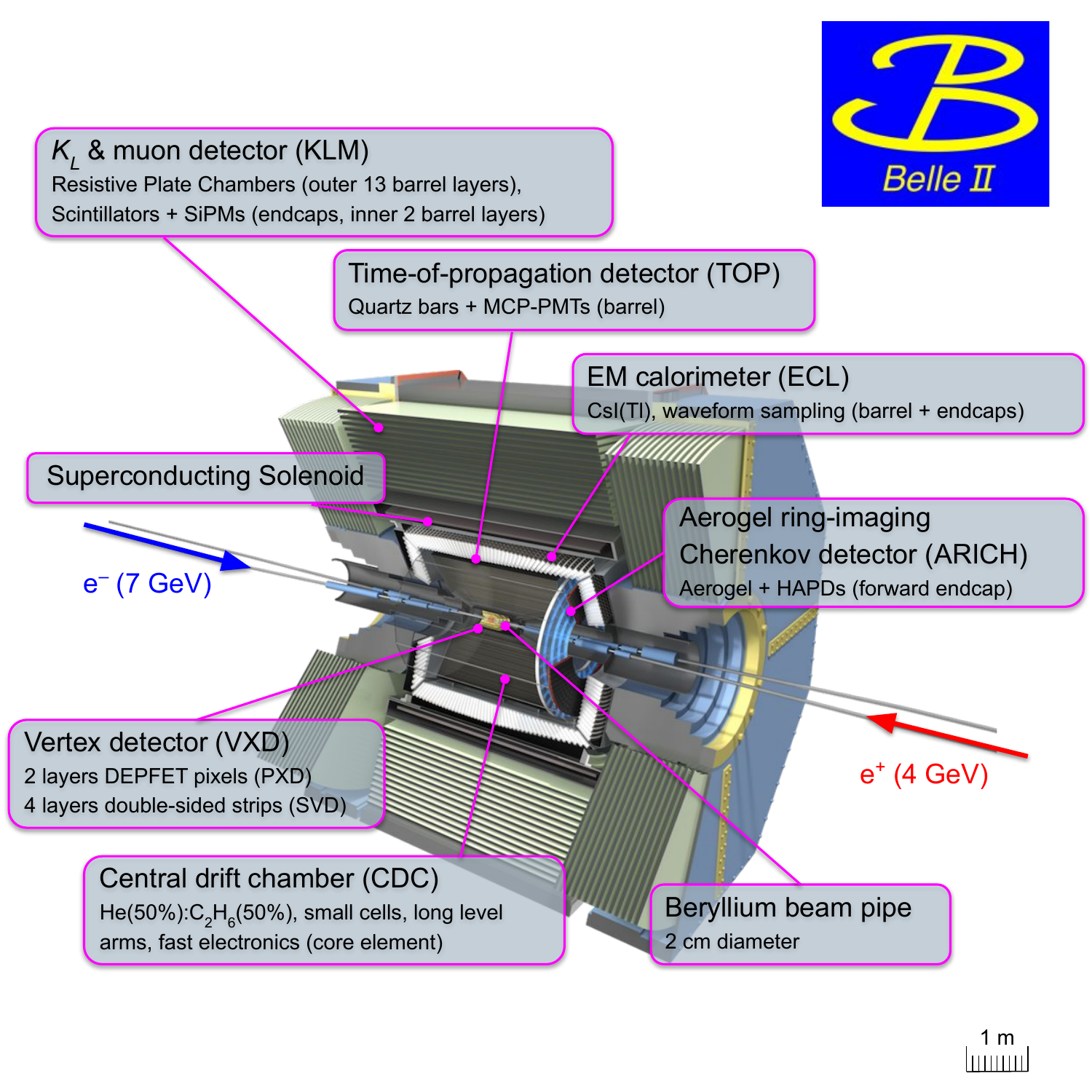}
\caption{\label{secIntro:fig4}Overview of the Belle~II detector.}

\end{figure}

\noindent {\it The Belle~II detector}, shown in Fig.~\ref{secIntro:fig4}, is a general-purpose particle spectrometer optimized for precise measurements of $B$-meson pairs via their decay products. The detector must maintain Belle's level of performance~\cite{Abashian2002,Brodzicka2012}, despite a reduced center of mass boost, and while operating in a much higher-background environment, which tends to reduce detector performance and longevity. Belle~II replaced a number of Belle sub-systems to satisfy this requirement and to have better vertexing and particle identification performance than Belle. Belle~II consists of several nested sub-detectors around the 1-cm radius beryllium beam pipe  surrounding the IP. 
The Belle~II sub-detector closest to the IP is the two-layer pixel detector (PXD). All 16 modules in the first PXD layer (L1), but only 4 out of the 24 modules in the second PXD layer (L2) have been installed to date. During LS1, we plan to install a new, fully assembled two-layer PXD, which will increase the detector's performance and tolerance of hit occupancy due to backgrounds~\cite{Forti2022}. The PXD is surrounded by four layers (L3-6) of the double-sided silicon strip vertex detector (SVD). Both PXD and SVD are surrounded by the CDC, which is filled with a $\rm He(\SI{50}{\%}) + C_{2}H_{6}(\SI{50}{\%})$ gas mixture. The CDC consists of 56~layers with 14336~sense wires of either axial or stereo orientation for precise measurements of charged particle trajectories. The charged-particle identification system is based on two sub-detectors: the barrel's TOP detector and the Aerogel Ring Imaging Cherenkov counter (ARICH) in the forward endcap region. The TOP is composed of 2-cm-thick quartz bars viewed by conventional and atomic layer deposition (ALD) types of MCP-PMTs, which are arranged into 16 readout slots. The ARICH consists of 4-cm-thick focusing aerogel radiators and 420 Hybrid Avalanche Photo Detectors (HAPDs), each having 144 readout channels. The HAPDs are grouped into 18 segments. For precise energy and timing measurements of particles, an electromagnetic calorimeter (ECL) is installed in the barrel and both endcaps. It is composed of 8736 CsI(Tl) crystals and is located inside a superconducting solenoid that provides a \SI{1.5}{T} magnetic field. Outside the magnet coil, a $K^{\rm 0}_{\rm L}$ and muon detector (KLM) is installed. The KLM has 12 and 14 scintillator strip layers read out by silicon photomultipliers in the forward (FWD) and backward (BWD) endcaps, respectively. The two innermost KLM barrel layers also utilize scintillators, while the remaining 13 barrel layers consist of glass-electrode resistive plate chambers (RPCs). A comprehensive overview of Belle~II upgrades planned for LS1 and LS2 can be found in Ref.~\cite{Forti2022}.

\subsection{Background types}

Belle~II hits generated by background shower particles deteriorate the detector’s physics performance. The radiation dose and neutron fluence from the background showers can also damage sensor components in the detector, such as silicon devices. Below we review the leading background components that are relevant at SuperKEKB.

\paragraph{Touschek background}
    
Touschek scattering~\cite{Piwinski1999}, one of the leading background sources at SuperKEKB, refers to Coulomb scattering between two particles in the same beam bunch. Such scattering causes the energies of the two scattered particles to deviate from the nominal beam energy, with one particle gaining, and the other losing, energy. The Touschek scattering rate is proportional to the beam current squared and inversely proportional to the number of bunches in the ring and the beam size. Due to the nano-beam scheme used at SuperKEKB, the beam size is much smaller than at KEKB, and consequently, the ring-integrated beam loss rate due to Touschek scattering is expected to be much higher. However, the Touschek loss rate inside Belle~II has been significantly suppressed by installing horizontal collimators near the IR.

\paragraph{Beam-gas background}

Beam-gas scattering by residual gas atoms in the beam pipe is another major background source at SuperKEKB. Beam-gas Coulomb scattering changes the direction of scattered beam particles, while beam-gas bremsstrahlung scattering reduces their energy. The beam-gas scattering rate is proportional to the 
residual gas pressure and to the beam current. The beam-gas Coulomb loss rate inside Belle~II is expected to be quite high due to the small diameter of the IP beam pipe and the extremely large vertical betatron function of the QCS. The loss rate in the detector has been greatly reduced by installing vertical collimators. However, the aperture of those collimators must be narrowed by moving their jaws towards the beam core, which can induce beam instabilities at high beam currents~\cite{Nakayama2012}.
    
\paragraph{Luminosity background}

Luminosity background is caused by beam collisions  at the IP. It is proportional to luminosity and expected to dominate at the target luminosity of SuperKEKB, which is about 30~times higher than the record of KEKB. 
    
One important luminosity background is from radiative Bhabha scattering ($e^+ e^- \rightarrow e^+ e^- \gamma$), where beam particles lose energy by emitting photons and therefore deviate from the nominal orbit. At KEKB, since a shared final focusing magnet scheme was employed, the outgoing beam orbits were off-center in the quadrupole magnets. Therefore, off-energy beam particles were strongly over-bent and easily lost inside the detector. Unlike KEKB, the final focusing magnets at SuperKEKB are separate for each ring, which relaxes the loss rate inside the detector. However, a small fraction of beam particles with large energy losses can still be lost inside the detector due to i) the strong magnetic field of the final focusing magnets, ii) intrinsic beam angular divergence at the IP, iii) angular diffusion by the radiative Bhabha process, iv) the kick from the detector solenoid field, and v) the leakage field from the other ring’s quadrupole magnets, especially for electrons as discussed in Ref.~\cite{OHUCHI2022165930}. At high luminosity, radiative Bhabha beam losses inside the detector dominate over other Belle~II backgrounds. 

Radiative Bhabha scattering can also give rise to neutron backgrounds incident upon Belle~II from the accelerator tunnel via the following mechanism: photons emitted in the radiative Bhabha process at the IP propagate along the beam axis and escape Belle~II. Such photons then hit accelerator magnets located \SIrange{10}{20}{m} downstream of the IP. Then, neutrons copiously produced via the giant photo-nuclear resonance~\cite{RevModPhys.47.713} scatter back towards the Belle~II detector. This background increases the hit occupancy in the outer  layers of the KLM. A dedicated study of this background component is described in Ref.~\cite{Schueler2021}.

In the two-photon process, $e^+ e^- \rightarrow e^+ e^- e^+ e^-$, beam particles lose energy by emitting low-momentum electron-positron pairs, and become a source of Belle~II background as described for the radiative Bhabha process. In addition, the emitted electron and positron curl in the Belle~II solenoid field. They can leave multiple hits in the inner Belle~II detectors if they have high enough transverse momentum. 
    
\paragraph{Synchrotron radiation background}

SR emitted from the beams is another source of background in the inner Belle~II detectors. Since the power of SR is proportional to the beam energy squared and the magnetic field strength squared, the HER electron beam is the main source of SR background. SR photons leave PXD and SVD hits with energy ranging from a few keV to several tens of keV. We pay special attention to this background because the inner layers of the SVD were severely damaged by HER SR in the early stages of the Belle  experiment. 
    
\paragraph{Injection background}
    
Since the beam lifetime of SuperKEKB is much shorter than an hour, top-up injections via a betatron injection scheme~\cite{Ohnishi2013} are performed during physics data taking. When the total beam current is below a set threshold ($\sim 99\%$ of the nominal beam current), charge is injected into buckets with low bunch-current, at a certain repetition rate (\SIrange{1}{25}{\hertz}). 
Newly injected bunches are perturbed and oscillate in the horizontal plane around the main stored beam. This causes increased background rates in Belle~II for a few milliseconds (ms) after injection each time when the newly injected bunch passes the IP. In order to avoid saturation of the readout, special trigger vetoes are applied, which lead to dead time in the data acquisition and, consequently, a reduction in recorded luminosity. A comprehensive description of the Belle~II trigger system is given in Ref.~\cite{Kou2019}. 

The amount and time structure of the injection background observed in Belle~II is shared online with the SuperKEKB operators and can be used to optimize the injection settings to keep backgrounds low. One of the most important and difficult tasks for SuperKEKB is maintaining stable injection background conditions for an extended period. 
    
\paragraph{Large beam loss accidents}

The accidental firing of one of the injection kicker magnets may perturb the stored beam during its  2-$\mathrm{\text\textmu s}$-long waveform towards a horizontal collimator, causing severe jaw damage.

Furthermore, for unknown reasons, the stored beam sometimes becomes unstable to the point of causing catastrophic or so-called \textit{sudden beam losses}. These losses have already caused several quenches of QCS magnets, damaged sensitive components of Belle~II, and significantly slowed down the planned luminosity increase. In other cases, the jaws of collimators were severely damaged, and beam operation was stopped for about a week to replace the jaws. Such events frequently occur when the beam current increases above \SI{0.5}{A}. A possible cause of these events is dust trapped in the beam pipe, but this is not yet fully understood. We are conducting detailed beam abort analysis using the timing information from the beam loss monitors installed around the ring. Such analysis may help us to identify the location where the initial beam loss occurred.

\subsection{\label{subsec:BackgroundMitigation}Background mitigation}
Here, we briefly review the crucial countermeasures against major, known background sources.

\paragraph{Collimators}

\begin{figure}[htbp]
\centering
\includegraphics[width=\linewidth]{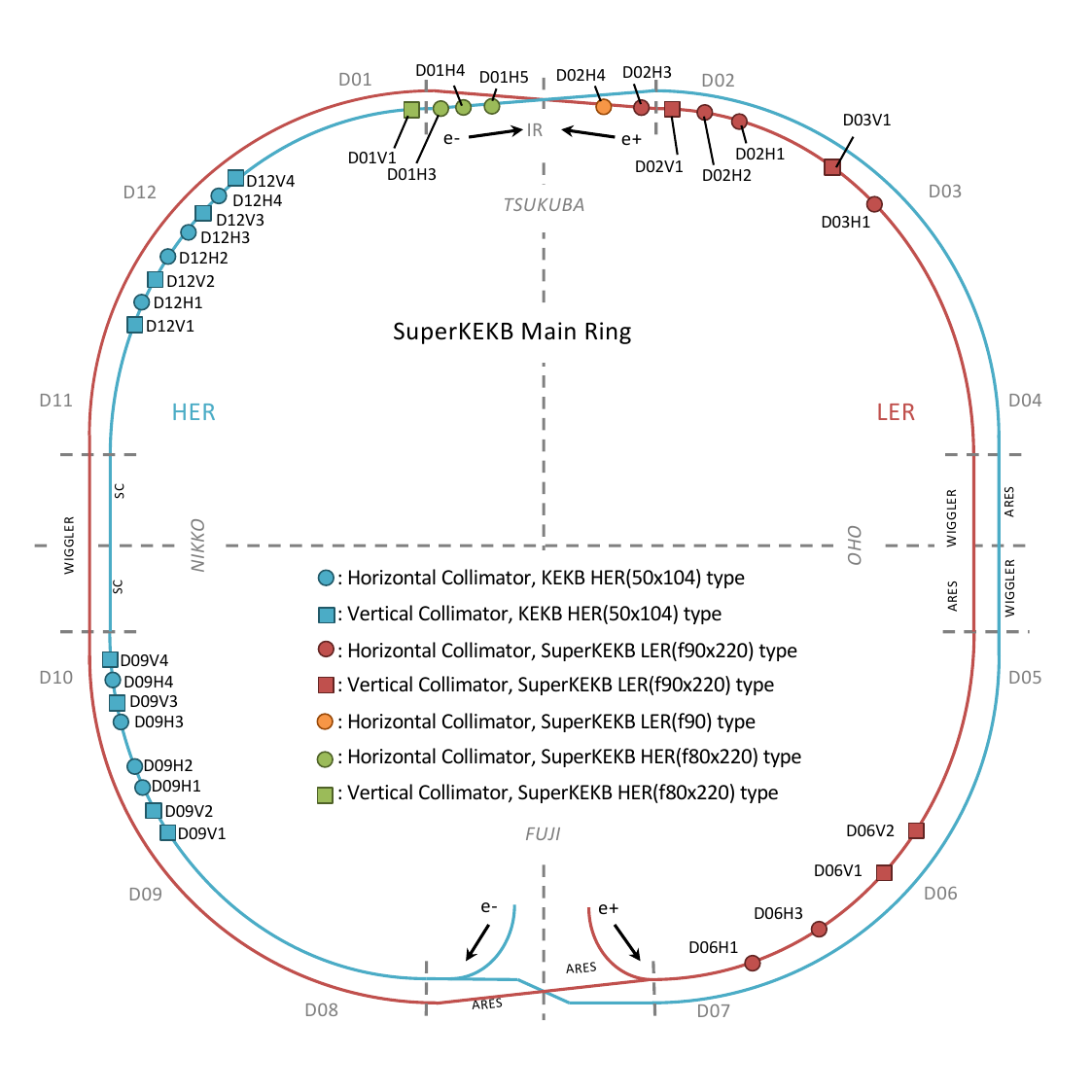}
\caption{\label{secIntro:fig1}Map of the SuperKEKB collimators used in 2021 and 2022. The letters V and H in the collimator names indicate vertical and horizontal movable jaws, respectively. There are twelve sections in each ring named D01 through D12.}

\end{figure}

Movable beam collimators are installed around SuperKEKB rings to stop beam particles with  large transverse deviation from the nominal beam orbit before they reach the IR and lead to background hits in Belle~II, see Fig.~\ref{secIntro:fig1}. Moreover, the collimators help protect Belle~II and the QCS magnets against large, unexpected beam losses, including those from accidental injection kicker firing. 

There are currently 11 collimators in the LER and 20 in the HER, see Fig.~\ref{secIntro:fig1}. There are two main types of collimators with different geometries: \textit{KEKB-type} collimators are asymmetric and have only one jaw, while \textit{SuperKEKB-type} collimators are symmetric with jaws on both sides. More details about the collimators can be found in Refs.~\cite{Natochii2021,Ishibashi2020}.

Horizontal collimators effectively stop Touschek scattered particles, while vertical collimators are mainly used to stop beam-gas Coulomb scattered particles. The vertical collimators must be closed to very small apertures of the order of \SI{1}{mm}, and therefore require a precise position control system. The small apertures can induce Transverse Mode Coupling Instabilities (TMCI) of the stored beam. They contribute to the overall machine impedance, resulting in an upper limit on the bunch current for stable operation~\cite{Chao1999}, 

\begin{equation}
    I_{\rm thresh.} = \frac{4 \pi f_{\rm s}E/e}{\sum\limits_{j}\beta_{j}k_{j}},
    \label{secIntro:eq1}
\end{equation}
where $I_{\rm thresh.}$ is the the bunch current threshold, $f_{\rm s}$ is equal to \SI{2.13}{kHz} and \SI{2.80}{kHz} for the LER and HER synchrotron frequency, respectively, $E$ is the beam energy, $e$ is the unit charge, and $\beta_{j}$ and $k_{j}$ are the beta function and kick factor~\cite{Natochii2021} of the $j$-th collimator, respectively.
In contrast, wide-open collimators increase beam losses in the IR, while too-narrow collimators reduce beam lifetime and injection efficiency. Therefore, each collimator should be set at the aperture that optimally balances backgrounds, lifetimes, injection performance and instabilities~\cite{Nakayama2012}. 

In our previous work on beam backgrounds~\cite{Natochii2021}, the simulation of the SuperKEKB collimation system was substantially improved, and it is now deemed reliable.
    
\paragraph{Detector shielding}
     
While collimators successfully reduce single-beam losses inside Belle~II, some fraction of stray beam particles still escape the collimators and are lost inside the detector. To protect the inner detectors from single-beam and luminosity background showers, tungsten shields are installed just outside the IP beam pipe and inside the vertex detector, but outside of the detector acceptance for physics signals. In addition, thick tungsten shields are also installed around the QCS, where the beam loss rate is estimated to be the highest due to a large betatron function.

\paragraph{IP beam pipe}
    
The IP beam pipe of SuperKEKB is carefully designed to reduce the SR background~\cite{Belle2TDR2010,Ohnishi2011}. SR from upstream of the IP is stopped by a tapered collimation part of the incoming pipe so that SR will not hit the central beryllium part of the IP beam pipe. Reflected SR will also not reach the central IP beam pipe, thanks to a ridge structure on the tapered surface of the incoming beam pipe. In addition, the effect of back-scattered SR is significantly reduced in SuperKEKB compared to KEKB. Because there is a separate QCS magnet for each ring, the outgoing beam orbit is almost straight and does not produce an SR fan. 

\subsection{Beam instrumentation relevant to background measurements}
This section lists the essential instrumentation (other than Belle~II) utilized to monitor beam parameters, the vacuum pressure in the beam pipe, and background levels at SuperKEKB.

\paragraph{Beam diagnostics}
    
In SuperKEKB, transverse beam sizes are measured by X-ray beam profile monitors (XRMs) and visible synchrotron radiation monitors (SRMs). For the analyses reported here, XRM data are used. The X-ray imaging system uses Cerium-doped yttrium-aluminum-garnet (YAG:Ce) scintillators combined with CMOS cameras~\cite{MULYANI20191}. A coded aperture imaging technique provides turn-by-turn vertical and horizontal beam size measurements with a spatial resolution of the order of $\SI{1}{\micro m}$ and $\SI{10}{\micro m}$, respectively~\cite{Mitsuka2022}. The bunch length is measured using a streak camera installed in each ring. Dedicated machine time is required to scan bunch lengthening from low ($\sim\SI{0}{mA/bunch}$) to high ($\sim\SI{1.4}{mA/bunch}$) bunch currents and to separate the lengthening due to single-beam effects, such as the longitudinal wakefield potential, from other influences, possibly from beam-beam interactions. Therefore, bunch length data are usually measured only once a year, to minimize interruptions of Belle~II data taking. Instantaneous and integrated luminosity measurements are provided by the Luminosity On-line Monitor (LOM), which is based on the rate of Bhabha scattering events measured by the ECL~\cite{Kovalenko2020}. At a  counting rate of about \SI{1}{Hz}, the system's statistic accuracy is 5\% at a luminosity of \SI{1e34}{cm^{-2}.s^{-1}} and the overall systematic uncertainty is estimated to be at the level of 1.7\%.
    
\paragraph{Vacuum system}
    
The vacuum system of the collider is designed to effectively mitigate i) higher order mode (HOM) power losses, ii) heat and gas loads due to the large SR power and photon density, and iii) the electron cloud and fast ion effects in the LER and HER, respectively. A distributed pumping system based on multilayer non-evaporable getter (NEG) strips~\cite{SUETSUGU2008153} is used to keep the vacuum pressure at the level of \SI{100}{nPa}, which is required to achieve  hours-long beam-gas lifetime. To measure the residual gas pressure in the collider beam pipe, about 300 cold cathode gauges (CCGs) are installed around each ring in roughly \SI{10}{m} intervals. These CCGs provide ultra-high vacuum pressure measurements above \SI{10}{nPa}. A dedicated 
vacuum pressure simulation shows that in the absence of circulating beams, the so-called \textit{base} gas pressure is almost the same at the center of the beam pipe as in the vicinity of the CCG. However, due to the finite conductance of the vacuum system, composed of the beam pipe, CCG, and vacuum pump, the \textit{dynamic} part of the pressure, which depends on the beam current and gas molecule desorption rate from the inner beam pipe walls, is approximately three times higher at the beam axis than at the CCGs; this factor was estimated from a simulation taking into account the conductance of the RF-shield screen between the beam channel and the pumping port and that of the pumping port itself~\cite{PhysRevAccelBeams.26.013201}. This factor of three, which is the same for both rings, is used in the beam-gas background study discussed later in the text.
    
SuperKEKB is instrumented with two residual gas analyzers (RGAs) in the D02 and D06 ring sections, see Fig.~\ref{secIntro:fig1}, to measure the molecular composition of residual gases in the beam pipe. These devices are mass spectrometers measuring mass-to-charge ratios of gas ion fragments. Due to the small number of RGAs, reliable information regarding the gas composition distribution around the collider is currently unavailable. Therefore, in simulation, an effective atomic number of $Z_{\rm eff.}=7$, mostly due to CO molecules, is assumed for the residual gas in the beam pipe ~\cite{Sahu1997}.
    
\paragraph{Background monitors}
    
Several dedicated detector systems are used to monitor backgrounds in the SuperKEKB IR and tunnel, and to ensure safe machine and detector operation. 
    
\begin{figure}[htbp]
\centering
\includegraphics[width=\linewidth]{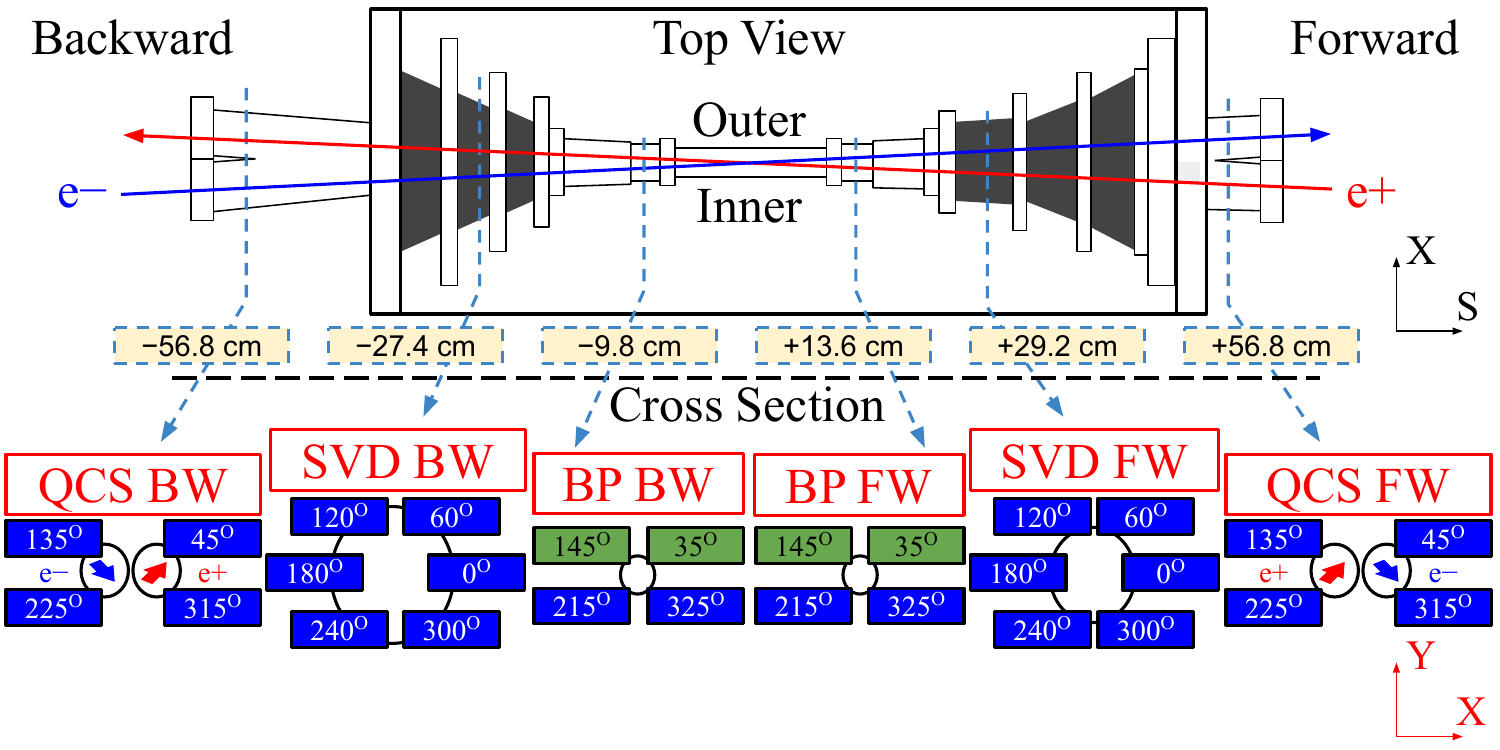}
\caption{\label{secIntro:fig2}Locations of diamond detectors in the IR. The dashed rectangles show the distance from the IP along the beam axis for each group of detectors. Azimuth angles of the detectors are indicated in rectangles. See text for further details.}

\end{figure}

\begin{itemize}
    \item Diamond sensor-based detectors (Diamonds)~\cite{Bacher2021} are used for radiation dose rate measurements in the IR, as shown in Fig.~\ref{secIntro:fig2}, where rectangles highlighted in blue and green indicate detectors used for dose rate monitoring at a \SI{10}{Hz} readout rate and fast beam abort monitoring at \SI{400}{kHz}, respectively. 
        
    \item The sCintillation Light And Waveform Sensors (CLAWS) detector system~\cite{Gabriel2021} is based on plastic scintillators and silicon photomultipliers. It monitors beam-induced backgrounds synchronized with the SuperKEKB injection. There are in total 32 CLAWS modules with 16 on the forward and 16 on the backward side of the IR around the final focusing magnets. The modules are located in four different longitudinal positions along the beam direction (approximately 1, 2, 3, and \SI{4}{m} from the IP) and four different azimuth angles (0$^{\circ}$, 90$^{\circ}$, 180$^{\circ}$ and 270$^{\circ}$) on each of the magnets.
        
    \item The BEAST TPC system uses six compact, high-resolution gaseous detectors~\cite{Jaegle:2019jpx} to provide directional and spectral measurements of the fast neutron flux~\cite{Schueler2021}. Currently, the detectors are located in the accelerator tunnel near Belle~II.
        
    \item Four \ce{^3He}~tube detectors~\cite{deJong2017}, installed around Belle~II, count thermal neutrons with kinetic energy below about \SI{0.025}{eV} through the following process:\\ \ce{^3_2He + ^1_0n \rightarrow ^3_1H + ^1_1H + \SI{764}{keV}}
        
    \item PIN photo-diodes~\cite{Ikeda2014} installed next to each collimator are used for fast beam loss monitoring around the movable jaws. 
        
    \item 5-m-long ion chambers~\cite{Ikeda2014} are mounted in cable racks on the outer wall along the accelerator tunnel. These air-filled gaseous detectors are used to measure beam losses. 
        
    \item New loss monitors, based on CsI-crystals with photo-multiplier tubes (PMTs) and electron-multiplier tubes (EMTs), were recently installed near SuperKEKB collimators. These new systems with good time synchronization capabilities are now used to pin down the location of sudden beam losses around the rings.
\end{itemize}
    
\paragraph{Beam abort system}
    
A dedicated fast beam abort system is used to dump unstable beams in order to avoid severe machine or detector damage. During commissioning Phases~1 and 2, the abort system included Diamonds (green rectangles, see Fig.~\ref{secIntro:fig2}), PIN photo-diodes and ionization chambers. In Phase~3, the system was augmented by including the four forward and four backward CLAWS detectors closest to the IP. These detectors can trigger a beam abort $\sim\SI{10}{\micro s}$ earlier than Diamonds, on average.

\section{\label{sec:CurrentBackgroundLevelsAndMargin}Current background levels and margin}

Here, we specify the main background vulnerabilities of each sub-detector. We also report on i) the current background rates seen during Belle~II operation in 2021, ii) the margin with respect to maximum acceptable rates, and iii) 
the recently observed detector performance degradation in 2022.
\subsection{PXD}

PXD is the inner-most detector, and its expected dominant background originates from the irreducible two-photon process where the low-momentum electron-positron pair spirals through the detector. Injection background and sudden beam losses are also of particular concern. The passing particles can deposit significant doses shortening the detector's lifetime and damage the detector by creating dead pixels or inefficient regions. As the closest detector to the IP, the PXD is also uniquely sensitive to the back-scatter of low-energy SR photons.

First, there is a limit on acceptable PXD occupancy due to bandwidth limitations. Assuming \SI{30}{kHz} trigger rate operation, some data loss will start to occur once the mean of the inner PXD layer occupancy exceeds 3\%. At 3\% occupancy, the offline performance will also degrade significantly because of cluster merging and an increased probability of associating wrong hits to tracks. 
Noticeable  degradation, however, starts  below this value. 

The PXD's second limit is associated with detector degradation due to radiation damage. A dose rate of \SI{2}{Mrad/smy}\footnote{The unit smy stands for a Snowmass year (\SI{1e7}{s}), which is the typical operation time of an accelerator facility.} for a 10-year-long operation of the device is deemed safe given the results from a dedicated X-ray irradiation campaign~\cite{SCHREECK2020163522}. Type inversion is not expected to occur before reaching a 1-MeV neutron equivalent fluence\footnote{The 1-MeV neutron equivalent fluence is the fluence of \SI{1}{MeV} neutrons producing the same damage in a detector material as produced by an arbitrary particle fluence with a specific energy distribution~\cite{Lemeilleur:1999cd,Casolaro:2020yeq}.} of \SI{1e14}{n_{eq}/cm^{2}}.

The current average PXD occupancy is below 0.3\%, suggesting PXD background levels are under control. At least once a year, however, significant beam losses have occurred, where 4--5\% of the so-called PXD switcher\footnote{The PXD switchers are the readout ASICs that switch on a pixel row to send the currents to the Drain Current Digitizers, which digitize the MOSFET currents from a row of pixels~\cite{Belle2TDR2010}.} channels were damaged. This makes the planned replacement of the current PXD with a new two-layer PXD during LS1 particularly valuable. 

Extrapolating current background levels to the predicted beam parameters before LS2 at the luminosity of \SI{2.8e35}{cm^{-2}.s^{-1}}, the PXD should be able to withstand the backgrounds and operate with a predicted average occupancy below 0.5\%, assuming 
the collimators can be operated close to ideal settings, and the total PXD background, including storage and injection components, stays below the detector limit.

\subsection{SVD}

In the SVD, the beam background increases the hit occupancy and causes radiation damage in the sensors. The increased hit occupancy, in turn, degrades the SVD tracking performance and increases data rates in the data acquisition system (DAQ). Radiation damage can affect the sensor leakage current, strip noise, and the full depletion voltage of sensors. It is important to estimate the expected SVD performance degradation over the entire lifetime of the experiment, given the expected background levels.

Radiation effects, respectively, from surface and bulk damage, are parameterized in terms of total ionizing dose released in the sensor (TID) and with non-ionizing energy loss (NIEL), expressed in a 1-MeV neutron equivalent fluence. Effects due to surface damage saturate after a relatively low integrated dose, about \SI{100}{krad}, while bulk effects are expected to dominate the SVD radiation damage in the long term.

The most restrictive limit on the SVD beam background levels is due to the degradation of the tracking performance, which limits the hit occupancy of the SVD inner-most layer (L3) to about 5\%, with a rejection of background hits based on the hit-time, that can be further refined. As for the integrated radiation damage, a deterioration of the SVD performance is expected after about \SI{6}{Mrad}, corresponding to about \SI{1.4e13}{n_{eq}/cm^{2}} of 1-MeV neutron equivalent fluence, due to a sizable reduction in the Signal-to-Noise.  After this level of irradiation, the increase in the sensor current, dominated by bulk damage, will produce noise from leakage current comparable to the one from the sensor capacitance, now dominant, thus increasing the noise 
by about $\sqrt{2}$. As for changes in the effective doping concentration and depletion voltage, no significant performance degradation is expected even after bulk type inversion and up to about \SI{2.5e13}{n_{eq}/cm^{2}}. 
This limit is based on the results of sensors used in the BaBar experiment, similar to the SVD ones, that were confirmed to be fully functional after irradiation  up to this level~\cite{Babardetector}.

Electrons and positrons are the dominant sources of beam background in the SVD, contributing to the hit occupancy and to radiation damage.  Neutrons and hadrons are the most effective for bulk damage, but it should be noted that electrons and positrons in the MeV-GeV energy range also contribute to bulk damage, although with a reduced effective cross-section for NIEL, properly accounted for in the conversion from particle fluence to 1-MeV neutron equivalent fluence. Electrons and positrons are either produced at the IP by the beam collisions or created off-IP by the scattering of the beam loss products in the accelerator components or the detector material and finally hitting the SVD. Neutrons are created off-IP and, although less abundant in the SVD, contribute via NIEL to the bulk radiation damage.

During operation in 2021, the hit occupancy averaged over the L3 sensors was 0.5\% at maximum, well below the occupancy limit of about 5\%. In the three-year operation of the SVD,  from 2019 to 2021, the first effects of radiation damage have been measured, consistent with  expectation, and with no degradation of the SVD performance~\cite{Uematsu:2655}.

The SVD is not always energized unlike the diamond sensors. Therefore, the integrated dose in the SVD is estimated from the dose measured by the diamond sensors on the beam pipe, and the measured correlation between the SVD occupancy and the diamond dose~\cite{Uematsu:2655, Massaccesi:2759}. The estimated integrated dose in the SVD L3 was about \SI{50}{krad} up to December~2021. The 1-MeV neutron equivalent fluence was evaluated to be about \SI{1.2e11}{n_{eq}/cm^{2}}, using a conversion factor from the integrated dose to the neutron equivalent fluence estimated by simulation.

Given the SVD limits of  about 5\% in L3 occupancy and about \SI{6}{Mrad} integrated dose, the SVD will be able to withstand, with a good safety margin, the background levels predicted before LS2 at the luminosity of \SI{2.8e35}{cm^{-2}.s^{-1}}, corresponding to about 1\% occupancy in L3 and  about \SI{70}{krad/smy}.


\subsection{CDC}

As the main tracking detector of Belle~II, a well-performing CDC is not only essential for tracking and the measurement of particle momenta but also for  trigger information and  particle identification via the measurement of specific ionisation in the chamber gas ($dE/dx$). Extra background hits caused in particular by LER Touschek and beam-gas scattering processes as well as by the injection background progressively degrade the CDC performance as the rate of background hits increases. 
The additional background hits can contaminate the physical signal of charged tracks, creating spurious tracks and smearing the kinematic variables of the reconstructed charged track.
Higher background levels also increase the overall current in the chamber,  increasing the risk of more rapid chamber ageing due to an accelerated buildup of deposits on the wires.
Finally, an increasing rate of single-event upsets (SEUs) in the front-end electronics of the CDC, caused by background neutrons with low kinetic energy, is another concern for the CDC operation~\cite{Higuchi:2012zz}. 
SEUs or other kinds of  CDC soft errors may stop the DAQ of Belle~II and decrease the data-taking efficiency. A planned upgrade of readout electronics during LS2 is expected to suppress the soft error rates.

The effect of background hits on the performance of the tracking algorithm has been studied with Monte-Carlo simulations~\cite{BERTACCHI2021107610}. To avoid degradation of the tracking performance, based on simulation at the luminosity of \SI{1.2e35}{cm^{-2}.{s}^{-1}}, a background hit rate of \SI{150}{kHz/wire} is acceptable, where the SVD stand-alone tracking retains high efficiency and CDC hits can be added to the SVD seed tracks. The CDC hit rates in 2021 for all layers were in the range from \SI{20}{kHz/wire} to \SI{50}{kHz/wire}, except for the first, inner-most layer with a hit rate of up to \SI{130}{kHz/wire}. The extrapolation of the background before LS2 at the luminosity of \SI{2.8e35}{cm^{-2}.s^{-1}} shows the CDC can run safely at beam currents up to $\sqrt{I_\mathrm{LER}I_\mathrm{HER}}\sim\SI{2.0}{A}$. The hit rates, except for the first layer, will reach \SIrange{50}{130}{kHz/wire} depending on the radial position of the layers, which is below the detector's limit. However, this simulation does not include the effect of the injection background during the trigger veto period, which leads to a strongly time-dependent overall chamber current.

\subsection{ARICH}

For the ARICH detector, there are three main adverse effects resulting from the beam-induced background. The first effect is neutron-induced silicon bulk damage in the avalanche-photo-diode chips (APDs) of the photon detectors (HAPDs), and the second is the Cherenkov photon background, mostly emitted by low-energy charged particles either in the aerogel or in the quartz window of the photon detectors. As a result of the first effect, the APD leakage current is steadily increasing with accumulated neutron fluence, eventually leading to a reduced Signal-to-Noise ratio and consequently either to the loss of photon detection efficiency or increased background hit rate. The increased background hit rate, resulting either from the APD noise or from the background Cherenkov photons, negatively impacts the ARICH particle identification performance. In neutron irradiation tests of HAPDs carried out prior to the ARICH construction, the leakage current remained tolerable ($<\SI{30}{\micro A/APD}$) at least up to a fluence of $\sim$\SI{1e12}{n_{eq}/cm^{2}}, which we consider as a conservative limit. The tolerable background photon hit rate was studied using the Monte-Carlo simulation, where we found the impact on performance to be negligible up to a hit rate of $\sim$\SI{1}{photon/HAPD/event}, where one event corresponds to \SI{250}{ns}. A third concern is that background neutron radiation is also a source of SEUs in the front-end electronics of the ARICH, which might, in some cases, break the DAQ chain and lower the data-taking efficiency.   

In the first three years of operation, from 2019 through 2021, the average increase in the APD leakage current was $\sim$\SI{0.3}{\micro A}, corresponding to a neutron fluence of $\sim$\SI{1e10}{n_{eq}/cm^{2}}, $\mathcal{O}(100)$ below the tolerable limit. The largest background photon hit rate observed so far was at the level of \SI{0.05}{photon/HAPD/event}, about a factor of 20 below the rate where performance will degrade noticeably. The rate of SEUs is at present observed to be about one per HAPD per day, and most SEUs are corrected on the fly 
in 
firmware \cite{Giordano2021}. In some cases, nonetheless, the DAQ  is halted. While such events are rare at present, further mitigation might have to be considered at increased SuperKEKB luminosity.     

\subsection{TOP}

The number of detected Cherenkov photons dictates the particle identification capability of the TOP. The typical number is \SIrange{20}{40}{photons/track}. To maintain good particle identification performance, it is essential to detect the  limited number of photons with high efficiency. However, one serious problem caused by the beam background in the TOP detector is a decrease in detection efficiency due to the degradation of the quantum efficiency (QE) of the MCP-PMTs. Measurements in our test bench showed that the QE degrades as a quadratic function of the accumulated output charge $\Sigma_Q$ of the MCP-PMT,
\[
\mathrm{Relative~QE} = 1-0.2\left(\frac{\Sigma_Q}{\tau_{\mathrm{QE}}}\right)^{2},
\]
where $\tau_{\mathrm{QE}}$ is the lifetime of the MCP-PMT, defined as the output charge corresponding to a relative QE of 0.8, compared to the initial value~\cite{MATSUOKA201793}. Three types of MCP-PMTs were installed, with the lifetime successfully extended during mass production of the MCP-PMTs.
The lifetime, measured in a test bench for a limited number of samples, is \SI{1.1}{C/cm^{2}} on average for the conventional type, \SI{10.5}{C/cm^{2}} on average for the ALD type and at least \SI{13.6}{C/cm^{2}} for the life-extended ALD type~\cite{MATSUOKA201793}. Degraded conventional and ALD MCP-PMTs will be replaced with the life-extended type during LS1 and LS2, respectively, for the TOP to withstand higher background rates.

The accumulated output charge is dominated by background Cherenkov photons from electrons and positrons generated when gamma rays hit the quartz bar, and Compton scatter or pair produce. To keep the accumulated output charge of the MCP-PMTs below the expected lifetime until their replacement or the end of Belle~II, we have imposed operational limits on the average MCP-PMT hit rate. The exact limit was updated from time to time based on QE projections. In 2021, the limit was \SI{3.0}{MHz/PMT}. The latest limit, in June~2022, was \SI{5.0}{MHz/PMT} for single-beam background, with an additional allowance for luminosity term, which cannot be mitigated by varying machine settings or collimators, and which scales with instantaneous luminosity as \SI{0.925}{MHz/PMT} per \SI{1e35}{cm^{-2}.s^{-1}}. The TOP MCP-PMT rate limit is the most stringent background limit among the Belle~II detector sub-systems, but has not limited accelerator operation with the typical average TOP background rate of about \SI{2}{MHz/PMT} in 2021.


In addition to the background hits in MCP-PMTs, we have observed that neutron backgrounds cause SEUs in the TOP front-end electronics boards. We have implemented an automated function to detect and correct the SEUs that occur in the configuration memory of programmable logic devices. Unfortunately, this function cannot correct errors that occur in bursts, as multiple simultaneous bit errors cannot be repaired. Such errors account for approximately 1\% of all detected errors. Furthermore, the function cannot detect SEUs that occur outside of configuration memory in the on-chip processor, as opposed to the programmable logic. Such errors occasionally occur in critical regions that can halt data taking until the front-end board is power cycled. In 2021 and 2022, manual interventions needed to recover such boards occurred at a rate of about 5~times a day, which was acceptable in terms of the active channel efficiency. However, a future rise in neutron backgrounds from higher beam currents could be a concern, as it would lead to more frequent halts  of the readout boards. 

\subsection{ECL}

The ECL detector is robust against backgrounds and does not have a hard background rate limit. However, its energy resolution slowly degrades as background rates increase. A dedicated ECL analysis is still in development.

\subsection{KLM}

The highest occupancy in the KLM occurs in the barrel's inner layers and the endcaps' outer layers. Although there is no significant difference between RPCs and scintillators in the current particle-identification performance, the scintillators are much more robust against backgrounds. The maximum rate limitations of KLM scintillators are being studied~\cite{Forti2022}. The long dead time of the RPCs during the recovery of the electric field after a discharge significantly reduces the detection efficiency under high background rates. Thus, this expected behavior was addressed in the design by instrumenting the two inner-most layers of the barrel and all layers of the endcaps with scintillators, while re-using RPCs from Belle for the 13 remaining barrel layers. The inner Belle~II sub-detectors effectively shield the inner KLM layers and reduce backgrounds produced inside the detector volume. Backgrounds originating outside Belle~II in the accelerator tunnel typically penetrate the outer KLM endcap layers first.

The most relevant background sources for the KLM are cosmic muons, fast neutrons produced by single-beam losses and radiative Bhabha scattering at low opening angles~\cite{Schueler2021}, and electronics noise. The spring 2021 background level of up to \SI{50}{Hz/cm^{2}} so far has not affected the performance of the KLM. It is planned to readout the signal waveform of the scintillators in the future to provide a higher-resolution ($<\SI{1}{ns}$) time measurement than is possible with the existing latch (binary) readout~\cite{Forti2022}. However, the new firmware will not be able to tolerate the occupancy observed in individual channels, especially for the outer endcap layers. A simpler readout mode for the affected region can be used to cope with this issue. Moreover, additional neutron shielding around Belle~II, planned for LS1, should suppress the flux of neutrons hitting the KLM and reduce the detector occupancy.

\subsection{Recent detector performance degradation}

In 2022, before the beginning of LS1, we gradually increased beam currents above \SI{1}{A} to reach a  luminosity higher than \SI{3e34}{cm^{-2}.s^{-1}}. In this period, several collimators were severely damaged by sudden beam losses, as introduced in Section~\ref{sec:Introduction}. Because beams incident on the damaged collimator jaw tips can lead to very high backgrounds, several damaged collimators had to be operated with wider apertures than optimal, resulting in a 
higher 
beam-induced backgrounds in Belle~II. This background increase caused noticeable reconstruction performance degradation in Belle~II, which is remarkable, as the rates, strictly speaking, were still well below the detector limits discussed above. The reduced performance in 2022 thus serves as a preview of the challenges Belle~II will face as luminosity and backgrounds increase, and highlights that despite careful simulation and component-level test-beam studies, there are likely to be a number of unanticipated detector-level, system-level, and software-level problems that only arise as backgrounds increase. The collaboration thus must remain vigilant and devote 
sufficient effort to understanding and mitigating backgrounds, as well as their impact on performance. Crucially, this must include background-level-dependent reconstruction algorithms and calibrations.

Here, we selectively mention some observations of performance degradation caused mainly by increased injection backgrounds due to damaged collimators and by injection chain imperfections at high beam currents. Although the direct impact of the injection background on the data acquisition is suppressed by applying a trigger veto in time with injections, the background can still lead to a noticeable performance degradation up to a few ms after the beam injection. This means that the background level becomes time-dependent, making this a good example of a situation where background-level-dependent reconstruction and calibration will be required. 

During 2022 the CDC gain dropped by about 15\% over the full detector volume. A drop in gain leads to less charge being collected and, consequently, fewer detected hits. The average number of CDC hits on high momentum tracks in di-muon events and on daughter tracks from $K_S^0$ was found to decrease by about 12\%. This decrease in the number of hits affects the momentum resolution for high-momentum tracks. The reduced collected charge and reduction in hits also lead to a reduction in particle identification performance via $dE/dx$, which only partly can be recovered by applying a more sophisticated calibration that takes into account the time of the event since the last injection. Possible causes for the reduced gain include a higher water content than expected in the CDC gas of inner layers, the increased voltage drop across a resistor in HV dividers, space charge effects of slowly moving ions, and faster than expected ageing of the CDC. The resistor mentioned will be replaced during LS1, but an overall quantitative understanding of the gain loss is still missing.

In the same running period, the ECL detector was also noticeably affected by the increased injection background, which is usually off-time and causes a CsI(Tl) crystal pedestal shift due to overlapping of the physics signal pulse ($\sim\SI{1}{\micro s}$) with neighboring background events. The shifted pedestal results in underestimated signal pulse height, decreasing the number of crystal hits and, consequently, less effective photon detection and electron identification.

Reduction of the injection background, and mitigation of its impact on performance, will be important tasks going forward. Close collaboration between SuperKEKB and Belle~II will be required.

\section{\label{sec:BackgroundSimulation}Background simulation}

This section provides a brief overview of the beam-induced background simulation in Belle~II. Reference~\cite{Natochii2021} provides a more comprehensive description of most of the Belle~II background simulation features implemented to date. A dedicated MC simulation is used to study beam loss processes in the machine, mitigate backgrounds, and evaluate the impact of the possible machine and detector upgrades on backgrounds. As introduced in Section~\ref{sec:Introduction}, the two dominant classes of beam backgrounds originating from the machine are i) single-beam backgrounds, from circulating charges in individual rings, and ii) luminosity backgrounds, from beam collisions. The simulation proceeds in two steps. First, we perform multi-turn tracking of electrons and positrons in the machine, collecting beam losses around each ring; then, we run simulations of the lost particle interactions with Belle~II to study the detector response to beam losses in the IR.

\subsection{Particle tracking in the machine}
The single-beam background simulation starts with the multi-turn particle tracking software framework Strategic Accelerator Design (SAD)~\cite{SAD2021}. SAD tracks scattered particles through a sequence of machine elements. Initialized with beam optics parameters and machine apertures, including collimators and beam pipes, SAD tracks particles for 1000 machine turns and collects beam losses around the ring. 

The tracking simulation starts by defining a set of $\sim$500 equidistant scattering regions around each ring, where bunches of particles are created. These particles are randomly generated within the 3D~volume of Gaussian bunches. The momentum and statistical weight of each particle are determined using well-known scattering theories: 
\begin{itemize}
    \item \textit{Coulomb scattering} is described by Rutherford's scattering formula, including a cutoff Coulomb potential and a screening effect for small angles~\cite{Chao1999,Jackson1962}.
    \item \textit{Bremsstrahlung} follows Bethe-Heitler's theory in Koch-Motz's description of complete screening in the Born approximation~\cite{Bethe1934,Koch1959}.
    \item \textit{Touschek scattering} is implemented through Moller's non-relativistic  differential cross section~\cite{Moller1932} using Bruck's formula~\cite{Bruck1974} for the loss rate calculation.
\end{itemize}
All tracked particles in SAD are scattered according to these processes. These stray particles are defined as lost if their spatial coordinates exceed the physical aperture of the machine.

\subsubsection{Collimators}
Collimators are the narrowest-aperture elements of the machine. They aim to protect sensitive elements of the accelerator and detector by absorbing the beam halo formed mainly by stray particles. Therefore, their accurate implementation into the particle tracking code is crucial.

\begin{figure}[htbp]
\centering
\includegraphics[width=\linewidth]{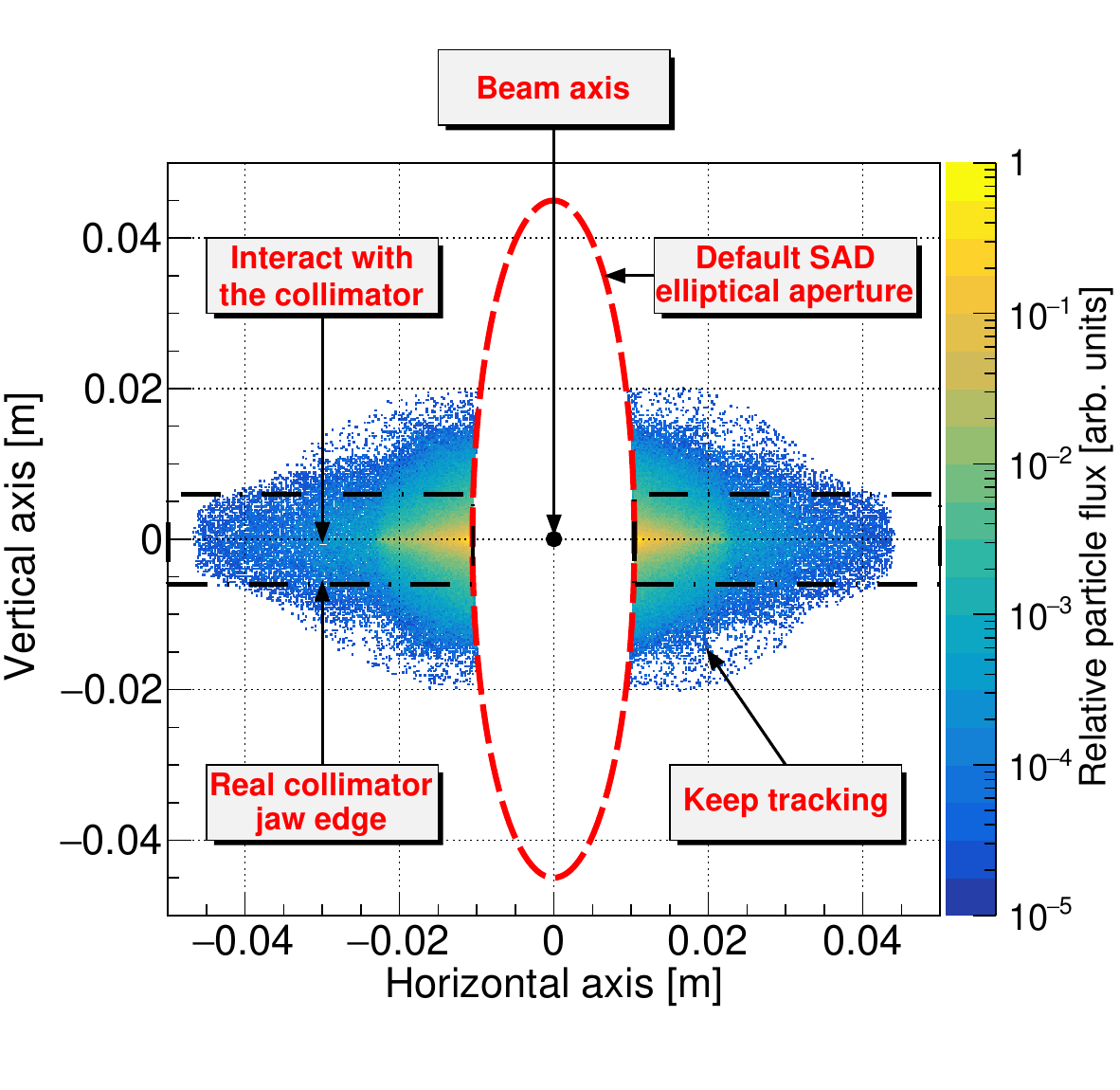}
\caption{\label{secB:fig1}Distribution of beam particles stopped by the LER D06H1 horizontal collimator (red dashed ellipse) in the original SAD simulation. The bin size is $\SI{0.2}{mm}\times\SI{0.2}{mm}$. Adapted from Ref.~\cite{Natochii2021}.}

\end{figure}

Recently, the simulation of the SuperKEKB collimation system description was significantly improved~\cite{Natochii2021} compared to Phase~1 and Phase~2 studies. By default in SAD, machine apertures, including collimators, are modeled as elliptical windows. Outside these windows, particles are considered as lost. This approximation is quite accurate for \textit{KEKB-type} collimators inherited from the KEKB collider. However, \textit{SuperKEKB-type} collimators have two opposite jaws with a rectangular shape and much thinner collimator heads ($\le \SI{10}{mm}$) along the beam axis. Therefore, a refined physical description of \textit{SuperKEKB-type} collimators, and a new simulation of the beam particle interaction with the collimator materials, was implemented. Figure~\ref{secB:fig1} shows the simulated distribution of beam particles lost at a horizontal collimator in the LER. The two black, dot-dashed rectangles show the newly implemented, more realistic collimator edge. Particles passing outside of the rectangular collimator jaws, labelled as ``Keep tracking'', are no longer (incorrectly) stopped by the simulated collimator, and instead remain in the simulated beam for tracking. This is a pivotal modification that substantially improved the simulation accuracy, as quantified by ratios between the experiment (Data) and simulation (MC), see next sections.

Moreover, for the background studies discussed in this paper, in addition to the introduced improvements in Ref.~\cite{Natochii2021}, we have recently implemented particle interaction with the copper collimator chamber. Although these improvements do not change the simulation results for the IR beam losses, they make our simulation code more reliable. 

\subsubsection{\label{subsec:PressureWeighting}Pressure weighting}

\begin{figure*}[htbp]
\centering
\includegraphics[width=\linewidth]{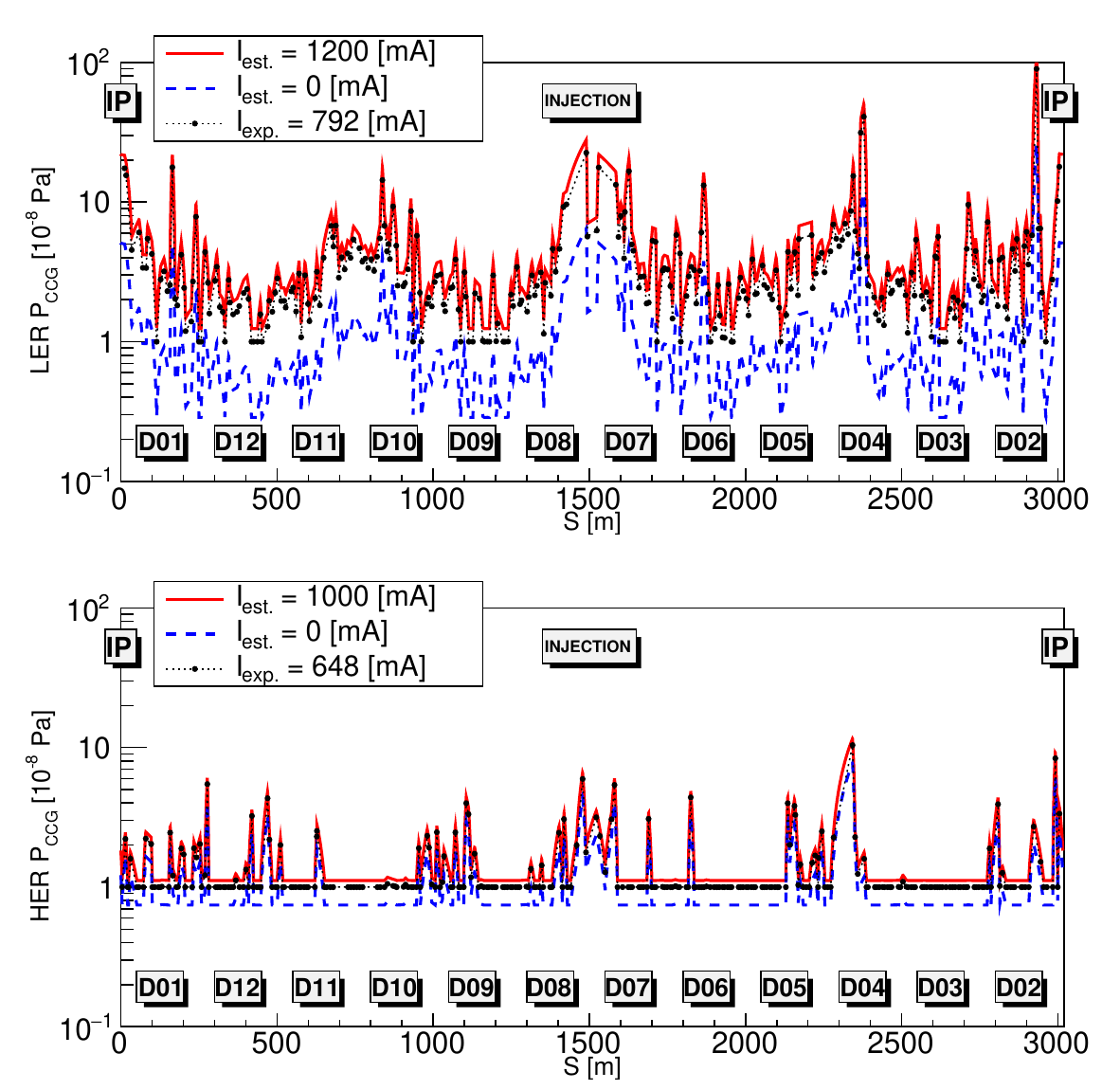}
\caption{Residual gas pressure versus longitudinal position in the LER (top) and HER (bottom). The black, dotted line with black data points ($I_\mathrm{exp.}$) shows pressure measured by CCGs in June~2021. The red, solid and blue, dashed lines ($I_\mathrm{est.}$) show estimated pressure at the beam currents listed in the legend. Labels identify different parts of the machine, such as the IP, the IR and the twelve sections of each ring, referred to as D01 through D12.}
\label{secB:fig2}
\end{figure*}

We describe an improved beam-gas background simulation, which was briefly mentioned in Ref.~\cite{Natochii2021}, and uses the measured residual gas pressure distribution. In the initial SAD simulation for Phase~1 and Phase~2 studies, we assumed a constant and uniform residual vacuum pressure of \SI{1}{nTorr} in both rings. However, the measured pressure depends both on position (Fig.~\ref{secB:fig2}) and time. Therefore, this paper uses the estimated gas pressure to re-weight lost particles depending on their scattering position.

When producing dedicated Belle~II Monte-Carlo samples for beam background studies, we typically use fixed reference beam currents ($I_\mathrm{LER} = \SI{1.2}{A}$, $I_\mathrm{HER} = \SI{1.0}{A}$) which are higher than those achieved during machine operation in 2020 and 2021 ($I \sim \SI{0.5}{A}$), but similar to those achieved in 2022. When the background simulation is validated against measurements, this is done at these reference currents. To facilitate the comparison for the beam-gas background, each lost particle after tracking in SAD is re-weighted by the measured gas pressure at its scattering location around the ring using the CCG gas pressure distribution measured at the time of studies, except that the distribution is initially re-scaled to the reference beam currents.

\begin{figure}[htbp]
\centering
\includegraphics[width=\linewidth]{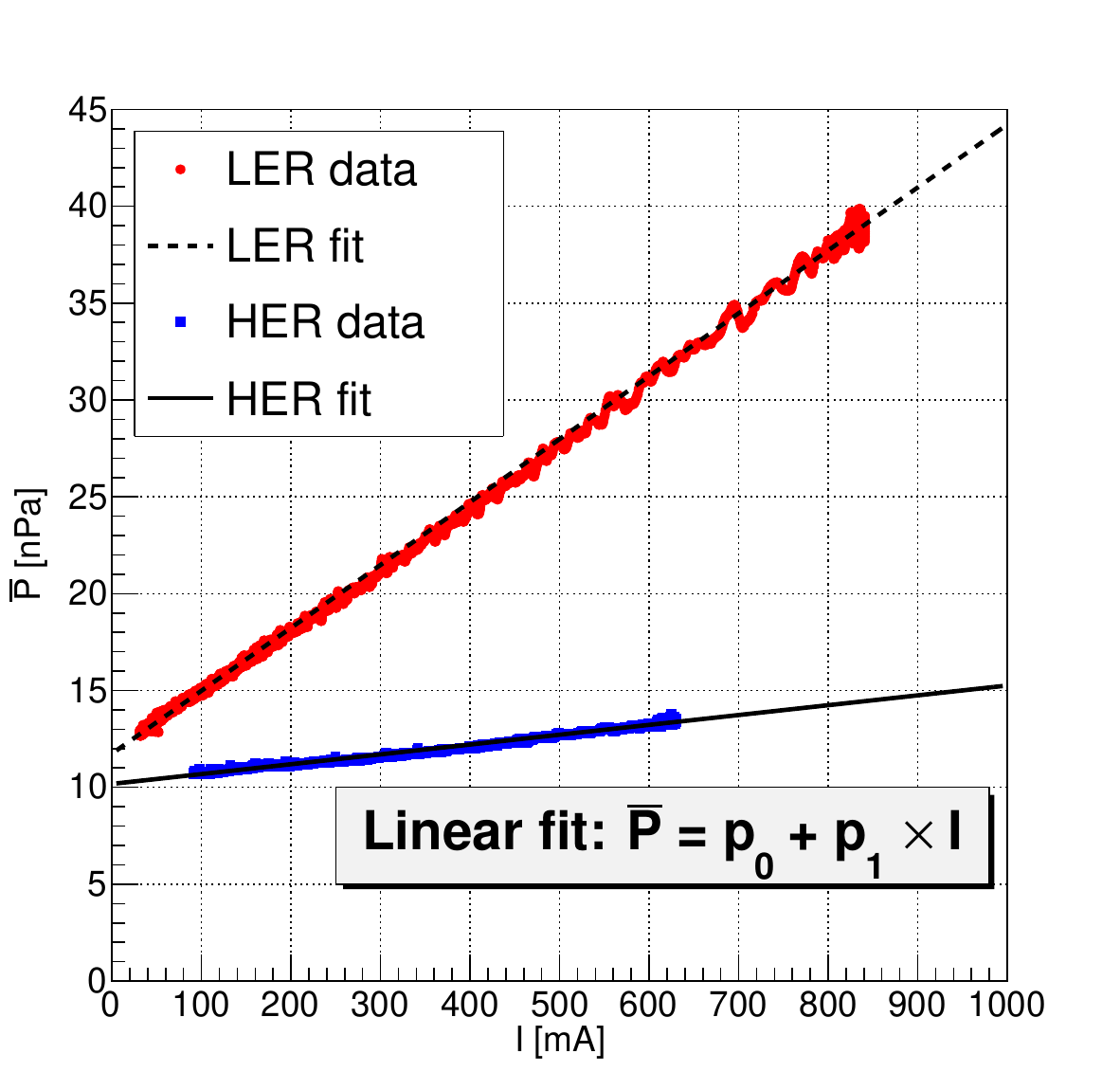}
\caption{\label{secB:fig3}Average ring gas pressure versus beam current measured in June~2021. 
}
\end{figure}

In order to re-scale the measured CCG gas pressure to the reference beam currents, we study the dependency between the averaged over-the-ring gas pressure as a function of the beam current. Figure~\ref{secB:fig3} shows the average ring pressure ($\bar{P}$) versus beam current ($I$) based on June~2021 CCG measurements. A linear fit, defined as $\bar{P} = p_\mathrm{0} + p_\mathrm{1} \times I$, determines i) the base pressure, $p_\mathrm{0} = \bar{P}(I = 0) = \bar{P}_\mathrm{0}$, which is the average ring pressure when there is no beam, and ii) the average dynamic pressure, $p_\mathrm{1} \times I = \mathrm{d}\bar{P}/\mathrm{d}I \times I$, where $\mathrm{d}\bar{P}/\mathrm{d}I$ is the average pressure increase per unit current, physically caused by gas molecules being released from the inner beam pipe walls. 
The obtained fit parameters are listed in Table~\ref{secB:tab1}.

\begin{table*}[htbp]
\centering
    \caption{\label{secB:tab1}Base ($p_\mathrm{0}$) and dynamic ($p_\mathrm{1}$) fit parameters of the measured gas pressure averaged over all CCGs as a function of beam currents, see Fig.~\ref{secB:fig3}.}
    \begin{tabular}{lcccc}
    \hline\hline
    \multirow{2}{*}{Date} &  \multicolumn{2}{c}{$p_\mathrm{0}$ [nPa]} & \multicolumn{2}{c}{$p_\mathrm{1}$ [nPa/A]}\\
     & LER & HER & LER & HER\\
    \hline
    May, 2020  & $14.77 \pm 0.01$ & $9.47 \pm 0.01$  & $52.08 \pm 0.06$ & $9.42 \pm 0.01$\\
    June, 2020 & $13.23 \pm 0.02$ & $9.34 \pm 0.01$  & $35.43 \pm 0.10$ & $8.51 \pm 0.02$\\
    June, 2021 & $11.74 \pm 0.01$ & $10.18 \pm 0.01$ & $32.48 \pm 0.01$ & $5.08 \pm 0.01$\\
    December, 2021 & $7.35 \pm 0.07$ & $9.13 \pm 0.01$ & $37.98 \pm 0.10$ & $5.16 \pm 0.01$\\
    \hline\hline
    \end{tabular}
\end{table*}

The measured vacuum pressure versus position is then re-scaled to the simulated beam currents (Fig.~\ref{secB:fig2}) as follows:
\begin{equation}
    P^\mathrm{est.}_\mathrm{CCG, i} = P^\mathrm{meas.}_\mathrm{CCG, i} \times \frac{p_\mathrm{0} + p_\mathrm{1} \times I}{\bar{P}^\mathrm{meas.}_\mathrm{CCG}},
    \label{secB:eq1}
\end{equation}
where $P^\mathrm{est.}_\mathrm{CCG, i}$ and $P^\mathrm{meas.}_\mathrm{CCG, i}$ are the estimated and measured gas pressure at the \textit{i}-th CCG, respectively, while $\bar{P}^\mathrm{meas.}_\mathrm{CCG}$ is the ring averaged pressure.

Although the sensitivities of the pressure gauges are limited to about \SI{1e-8}{Pa}, the scaling helps estimate the pressure below that limit at $I = \SI{0}{A}$. Moreover, for the ring-averaged gas pressure calculation in Fig.~\ref{secB:fig3}, we consider the saturated value (\SI{1e-8}{Pa}) as a real measured pressure at the given beam current. Therefore, this assumption leads to overestimated base and underestimated dynamic average pressure in the HER. The peaky, non-uniform distribution of the residual gas pressure in Fig.~\ref{secB:fig2} results in an unequal contribution of lost particles to beam losses depending on their scattered location around the ring.

\subsection{Particle interactions with the detector}

\begin{figure}[htbp]
\centering
\includegraphics[width=\linewidth]{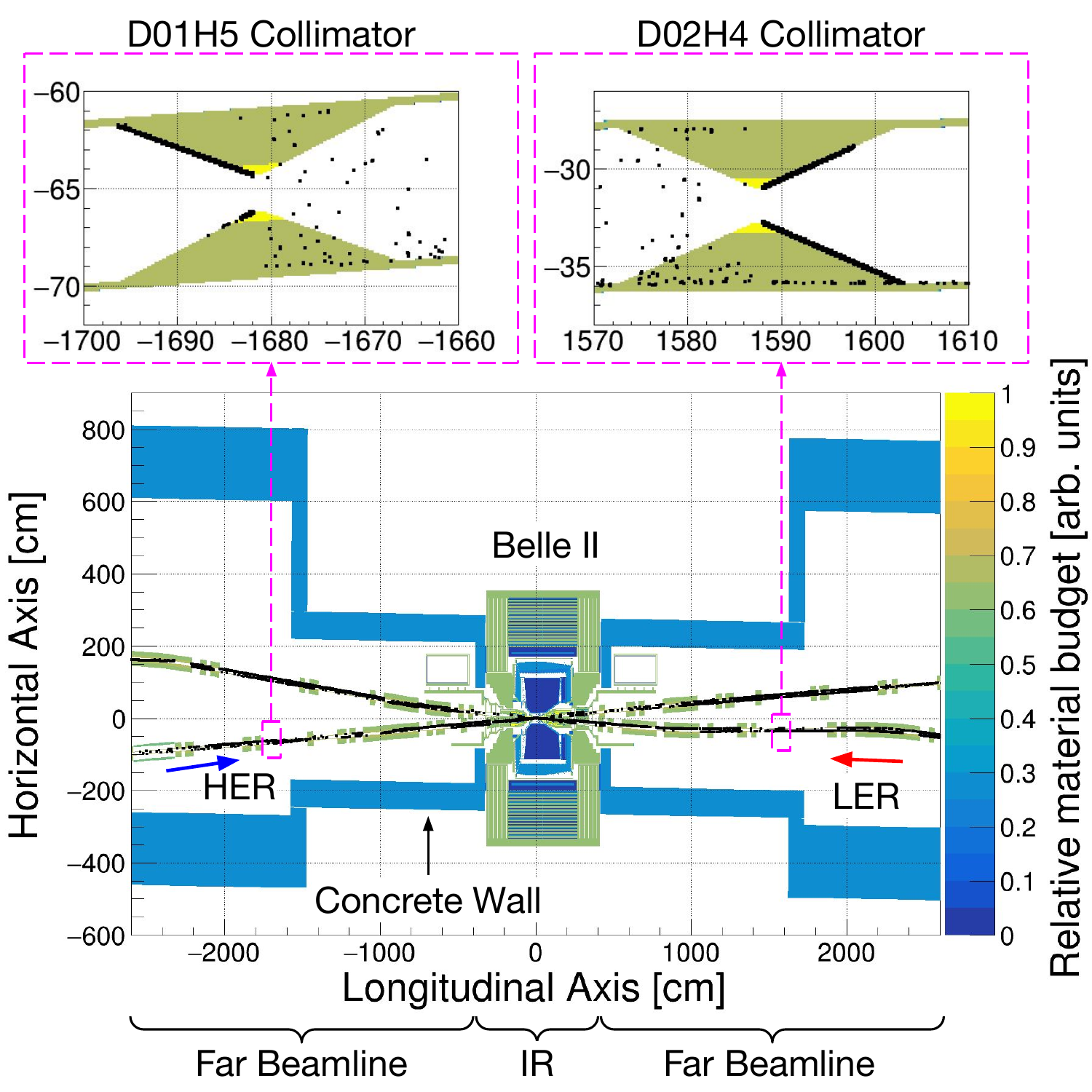}
\caption{\label{secB:fig4}Simulated beam losses on internal surfaces of beam pipe walls. Two top figures show beam loss distribution on upstream surfaces of horizontal collimators.}

\end{figure}

We use the Geant4 (v10.6.3) toolkit~\cite{Agostinelli2003,Allison2006,Allison2016} embedded into the Belle~II Analysis Software Framework (basf2)~\cite{Kuhr:2018lps,basf2-zenodo} to simulate the detector response to beam-induced backgrounds using the FTFP\_BERT\_HP Geant4 physics list~\cite{geant4-phys-list}. Beam-gas and Touschek scattered particles lost near Belle~II in SAD are passed from SAD to Geant4 at the inner surface of beam pipes and collimators. We have recently improved the SAD to Geant4 interface in order to accurately account for the curvature of beam pipes and the tapered shapes of collimators. The Geant4-simulated region extends out longitudinally $\sim\SI{30}{m}$ on both sides of the IP. The geometry consists of the IR ($\pm\SI{4}{m}$), where Belle~II is located, and the so-called \textit{far beamline region}, immediately outside the IR, where the Geant4 geometry includes elements such as magnets, beam pipes, tunnel walls, collimators, and shielding, see Fig.~\ref{secB:fig4}. We invested much effort in improving the IR and far beamline geometry description in Geant4. This has made our simulation more consistent with measurements and hence more reliable.

The luminosity background is simulated using dedicated event generators, followed by Geant4, and the same geometry as described above. SAD is not required in this case. Luminosity backgrounds considered include radiative Bhabha and two-photon processes (Section~\ref{sec:Introduction}). The specific event generators used are BBBREM~\cite{KLEISS1994372} and BHWIDE~\cite{JADACH1997298} for small ($<\!0.5^\circ$) and big ($>\!0.5^\circ$) scattering angle radiative Bhabha processes, respectively, and AAFH~\cite{BERENDS1985441} for two-photon processes. 

At the end of the simulation, we collect detector hits for each sub-system of Belle~II and compare the simulated against measured background observables.

\section{\label{sec:MeasuredBackgroundComposition}Background decomposition procedure}

Here, we give an overview of how the beam-induced background composition at SuperKEKB is measured and modeled. Table~\ref{secC:tab3} lists sub-detector elements and related background observables used for the analysis.

\begin{table}[htbp]
\centering
    \caption{\label{secC:tab3}Belle~II background observables. The twelve diamond detectors (4~QCS-FWD, 4~QCS-BWD, 4~BP) are shown as blue rectangles in Fig.~\ref{secIntro:fig2}.}
    \begin{tabular}{lccc}
    \hline\hline
    Sub-detector & Element & Observable & Units \\
    \hline
    Diamonds & 12 detectors & Dose rate & mrad/s \\
    PXD & 40 modules & Occupancy & \% \\
    SVD & 4 layers & Occupancy & \% \\
    CDC & 56 layers & Hit rate & kHz/wire \\
    TOP & 16 slots & Hit rate & MHz/PMT \\
    ARICH & 18 segments & Photon rate & MHz/HAPD \\
    KLM & 41 layers & Hit rate & MHz/layer \\
    \hline\hline
    \end{tabular}
\end{table}

\subsection{Background models}
\subsubsection{\label{subsec:SingleBeam}Single-beam}
In Belle~II, the two main single-beam background components are due to beam-gas and Touschek scattering of circulating charges in the vacuum beam pipe. To disentangle these two sources of particle losses, we employ a so-called \textit{heuristic} model, which was first introduced in Phase~1~\cite{LEWIS201969}, improved in Phase~2~\cite{Liptak2021}, and further refined here. Following the beam-gas and Touschek scattering theories~\cite{Ohnishi2013,Chao1999,Wolski2014}, we model measured observables, largely Belle~II detector rates and occupancies (Table~\ref{secC:tab3}), as

\begin{equation}
    \mathcal{O}_\mathrm{beam-gas} = B \times I\bar{P}_\mathrm{eff.},
    \label{secC:eq1}
\end{equation}

\begin{equation}
    \mathcal{O}_\mathrm{Touschek} = T \times \frac{I^{2}}{n_\mathrm{b}\sigma_\mathrm{x}\sigma_\mathrm{y}\sigma_\mathrm{z}},
    \label{secC:eq2}
\end{equation}
where $\mathcal{O}_\mathrm{beam-gas}$ ($\mathcal{O}_\mathrm{Touschek}$) is the beam-gas (Touschek) component; $B$ ($T$) is the beam-gas (Touschek) sensitivity, $I$, $\bar{P}_\mathrm{eff.}$ and $n_\mathrm{b}$ are the beam current, the ring average effective residual gas pressure seen by the beam, and the number of bunches in each ring, respectively. The bunch volume is defined by the product of $\sigma_\mathrm{x}, \sigma_\mathrm{y}$, and $\sigma_\mathrm{z}$, which are bunch sizes in the XY-plane and bunch length along the beam axis, respectively.

While the transverse bunch sizes are measured continuously during background studies, the longitudinal bunch length is not. Therefore, we instead parameterize the bunch length dependence on other beam parameters. Measurements of this dependence in commissioning Phase~3 are discussed further in Ref.~\cite{Mitsuka2020}. For our analysis, we use updated results~\cite{Ikeda2021} performed in 2020 and 2021 for the HER and LER, respectively. In our fit model, the bunch length is parameterized as follows:

\begin{equation}
    \sigma_\mathrm{z}^\mathrm{LER}[\mathrm{mm}] = 5.4466 + 1.7642 \times \frac{I^\mathrm{LER}[\mathrm{mA}]}{n_\mathrm{b}^\mathrm{LER}},
    \label{secC:eq8}
\end{equation}

\begin{equation}
    \sigma_\mathrm{z}^\mathrm{HER}[\mathrm{mm}] = 6.0211 + 1.3711 \times \frac{I^\mathrm{HER}[\mathrm{mA}]}{n_\mathrm{b}^\mathrm{HER}}.
    \label{secC:eq9}
\end{equation}

During machine operation, there is a constant flow of desorbed gas from the beam pipe to the vacuum pumps. As a result of this flow, the finite conductance of the vacuum system and the location of the CCGs, the ring average pressure at center of the beam pipe, $\bar{P}_\mathrm{eff.}$, which is the pressure relevant for beam-gas scattering, is higher than the pressure measured by CCGs. We use the CCG data to estimate $\bar{P}_\mathrm{eff.}$. It is assumed, based on geometry, that the dynamic pressure measured by CCGs, $I(\mathrm{d}\bar{P}/\mathrm{d}I)_\mathrm{CCG}$, is three times lower than at the center of the beam pipe, while the base pressure, $\bar{P}_\mathrm{0, CCG}$, is assumed to be the same as seen by the beam. Therefore, $\bar{P}_\mathrm{eff.}$ can be obtained from the measured CCG gas pressure averaged over the ring as follows

\begin{equation}
    \bar{P}_\mathrm{eff.} = 3 I(\mathrm{d}\bar{P}/\mathrm{d}I)_\mathrm{CCG} + \bar{P}_\mathrm{0, CCG} = 3 \bar{P}_\mathrm{CCG} - 2 \bar{P}_\mathrm{0, CCG},
    \label{secC:eq3}
\end{equation}
where $\bar{P}_\mathrm{CCG} = I(\mathrm{d}\bar{P}/\mathrm{d}I)_\mathrm{CCG} + \bar{P}_\mathrm{0, CCG}$ as discussed in Section~\ref{subsec:PressureWeighting}. Table~\ref{secC:tab2} lists extrapolation parameters of $\bar{P}_\mathrm{eff.}$ as a function of beam currents for so-called \textit{sensing} ring sections, where the measured CCG pressure averaged over the ring section behaves linearly along the full range of the measured beam current (\SIrange{10}{1000}{mA}) above the CCG hardware limit of \SI{10}{nPa}. In Table~\ref{secC:tab2}, the averaging over the ring before fitting is done as an arithmetic mean over the ring sections specified in the second and third columns for the LER and HER, respectively. The parameters are used for Data/MC calculation, where $\bar{P}_\mathrm{eff.}$ is extrapolated towards simulated beam currents, see later in the text.

\begin{table*}[htbp]
\centering
    \caption{\label{secC:tab2}Base ($\bar{P}_\mathrm{0, CCG}$) and dynamic ($(\mathrm{d}\bar{P}/\mathrm{d}I)_\mathrm{CCG}$) fit parameters of the measured CCG gas pressure averaged over sensing ring sections as a function of beam currents.}
    \begin{tabular}{lcccccc}
    \hline\hline
    \multirow{2}{*}{Date} & \multicolumn{2}{c}{Sensing ring sections} & \multicolumn{2}{c}{$\bar{P}_\mathrm{0, CCG}$ [nPa]} & \multicolumn{2}{c}{$(\mathrm{d}\bar{P}/\mathrm{d}I)_\mathrm{CCG}$ [nPa/A]}\\
     & LER & HER & LER & HER & LER & HER\\
    \hline
    May, 2020  & D01-D12 & D02, D04, D09 & $14.79 \pm 0.22$ & $9.66 \pm 0.58$  & $52.08 \pm 1.25$ & $11.54 \pm 1.44$\\
    June, 2020 & D01-D12 & D02, D04, D09 & $13.07 \pm 0.44$ & $10.13 \pm 0.79$  & $36.23 \pm 2.00$ & $9.77 \pm 2.04$\\
    June, 2021 & D01-D11 & D02, D04, D09, D12 & $12.68 \pm 0.16$ & $10.72 \pm 0.04$ & $30.55 \pm 0.57$ & $6.24 \pm 0.08$\\
    December, 2021 & D01-D11 & D02, D04, D12 & $7.92 \pm 0.95$ & $10.52 \pm 0.03$ & $39.76 \pm 1.42$ & $5.40 \pm 0.04$\\
    \hline\hline
    \end{tabular}
\end{table*}

The overall single-beam background observable for each ring is defined as a sum of beam-gas ($\mathcal{O}_\mathrm{beam-gas}$) and Touschek ($\mathcal{O}_\mathrm{Touschek}$) components plus a constant pedestal ($D$) which represents the detector electronics noise or calibration offset
\begin{equation}
    \mathcal{O}_\mathrm{single} = B \times I\bar{P}_\mathrm{eff.} + T \times \frac{I^{2}}{n_\mathrm{b}\sigma_\mathrm{x}\sigma_\mathrm{y}\sigma_\mathrm{z}} + D,
    \label{secC:eq5}
\end{equation}
where $\bar{P}_\mathrm{eff.}$ is defined in Eq.~\eqref{secC:eq3} with $\bar{P}_\mathrm{CCG}$ calculated as an average CCG gas pressure over sensing ring sections, and $\bar{P}_\mathrm{0, CCG}$ taken from Table~\ref{secC:tab2}, assuming the base pressure stays stable during the study.

During the early stage of the commissioning Phase~3, a large photon background was observed for some runs in a few modules of the PXD detector. Since the interaction region is designed so that no direct SR photons hit the central beam pipe, most of the SR background consists of secondary photons. To account for the SR background in our model, $\mathcal{O}_\mathrm{SR}$, which is proportional to the HER beam current, we extend the HER heuristic fit formula for the PXD detector as follows

\begin{equation}
    \mathcal{O}_\mathrm{single}^\mathrm{PXD} = \mathcal{O}_\mathrm{single} + S \times I,
    \label{secC:eq10}
\end{equation}
where $S$ is the SR sensitivity.

\subsubsection{Luminosity}
The luminosity background is, by definition, linearly proportional to the instantaneous luminosity ($\mathcal{L}$). We describe this background component as follows

\begin{equation}
    \mathcal{O}_\mathrm{lumi} = L \times \mathcal{L},
    \label{secC:eq6}
\end{equation}
where $L$ is the luminosity sensitivity. The luminosity background can be evaluated from measured observables, $\mathcal{O}_\mathrm{meas.}$, during collisions by subtracting single-beam backgrounds from non-injection data:

\begin{equation}
    \begin{split}
        \mathcal{O}_\mathrm{lumi} = \mathcal{O}_\mathrm{meas.} \\
        & - (B \times I\bar{P}_\mathrm{eff.} + T \times \frac{I^{2}}{n_\mathrm{b}\sigma_\mathrm{x}\sigma_\mathrm{y}\sigma_\mathrm{z}})^\mathrm{LER}\\
        & - (B \times I\bar{P}_\mathrm{eff.} + T \times \frac{I^{2}}{n_\mathrm{b}\sigma_\mathrm{x}\sigma_\mathrm{y}\sigma_\mathrm{z}})^\mathrm{HER}\\
        & - \frac{1}{2}(D^\mathrm{LER} + D^\mathrm{HER}).
    \label{secC:eq13}
    \end{split}
\end{equation}

Note that for each individual sub-detector element, there are specific observables listed in Table~\ref{secC:tab3} and sensitivities: $B^\mathrm{LER,HER}$, $T^\mathrm{LER,HER}$, $D^\mathrm{LER,HER}$, and $L$, plus $S^\mathrm{HER}$ for the PXD SR background.

\subsection{\label{subsec:DedicatedBgStudy}Dedicated background studies}

Approximately twice a year, the Belle~II beam background group performs dedicated beam-induced background measurements at SuperKEKB. The major goals are to investigate the background composition and to compare measurements against simulation. This information is needed to make reliable projections of future backgrounds and to perform targeted background mitigation. We focus on four comprehensive studies under stable and well-controlled machine conditions, which were conducted on May~9 ($\beta^{*}_\mathrm{y}$ = \SI{1.0}{mm}) and June~27 ($\beta^{*}_\mathrm{y}$ = \SI{0.8}{mm}) in 2020, and June~16 ($\beta^{*}_\mathrm{y}$ = \SI{1.0}{mm}) and December~20 ($\beta^{*}_\mathrm{y}$ = \SI{1.0}{mm}) in 2021.  

\begin{figure*}[htbp]
\centering
\includegraphics[width=\linewidth]{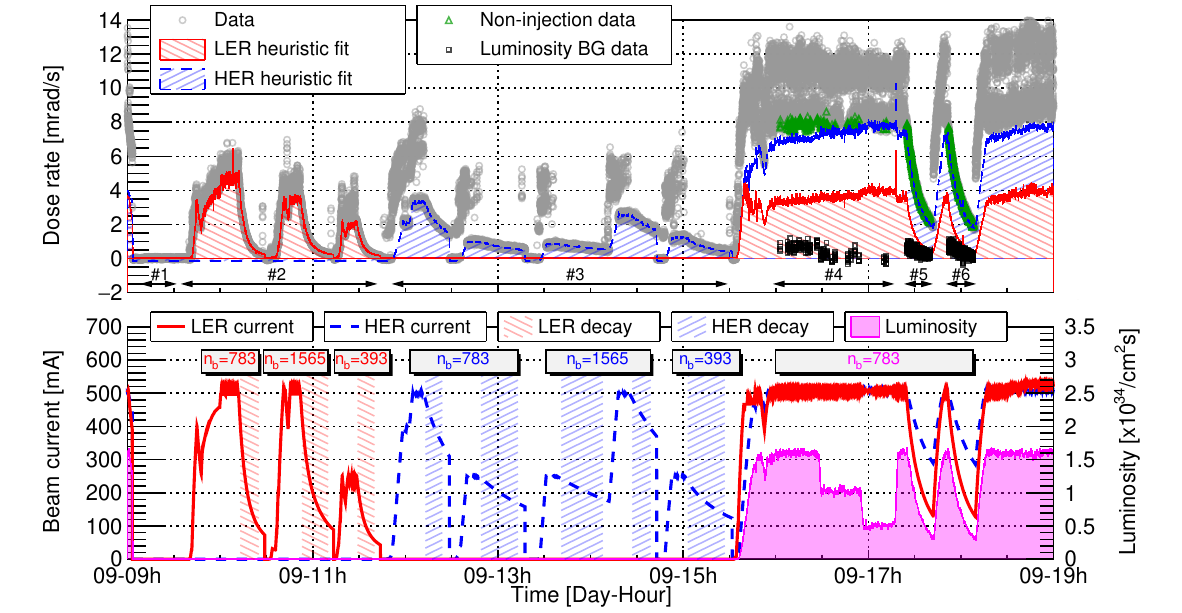}
\caption{\label{secC:fig1}Example of dedicated beam background measurements on May~9, 2020. Top: BP-FW-325 diamond detector dose rate; bottom: SuperKEKB machine parameters. See text for detailed discussion.}

\end{figure*}

Figure~\ref{secC:fig1} illustrates the study performed on May~9, 2020. The top plot shows an example of one background observable, a measured diamond detector dose rate (open gray circles). The bottom plot shows measured beam parameters. The study consists of three types of measurements identified in the top plot: i) no-beam (\#1), to estimate statistical fluctuation of the measured observable without beams circulating in the machine; ii) single-beam (\#2 LER, \#3 HER), where one ring at a time is filled with a beam of particles; iii) luminosity (\#4-6), to study beam losses during collisions of the two beams. For the single-beam background measurements, we inject only one beam to a current of $\sim \SI{0.5}{A}$ and collect data during about \SI{5}{min} of top-up injections. This allows the gas pressure to settle and provides data for the study of the injection background. Then, the beam current is left to decay for about \SI{15}{min} with no injection. This data sample is defined as \textit{beam decay} and shown as hatched bands in Fig.~\ref{secC:fig1} (bottom). Varying the number of bunches in the ring allows us to disentangle the beam-gas and Touschek components, as only the latter depends on the number of bunches at fixed beam current, see Eq.~\eqref{secC:eq2}. We use Eq.~\eqref{secC:eq5} to fit measured observables during the single-beam study for each ring separately, which yields background sensitivities for the beam-gas ($B^\mathrm{LER,HER}$) and Touschek ($T^\mathrm{LER,HER}$) components, as well as detector pedestal ($D^\mathrm{LER,HER}$). The single-beam fit results, see hatched areas in Fig.~\ref{secC:fig1} (top, where the LER and HER extrapolated backgrounds are shown as stacked histograms), can then be extrapolated to other times using machine parameters and are used in the following luminosity background measurements. To study luminosity backgrounds for a fixed number of bunches, we 1) scan the luminosity during top-up injection for both beams at nominal currents (\#4) by applying a vertical orbit offset between the colliding beams, and then 2) stop injection, leaving both beams to decay (\#5 and \#6). The luminosity background (open black squares) is calculated as the difference between non-injection data (open green triangles) collected during periods 1) and 2) and the sum of the extrapolated LER and HER single-beam heuristic fits, see Fig.~\ref{secC:fig1}. 

\begin{figure}[htbp]
\centering
\includegraphics[width=\linewidth]{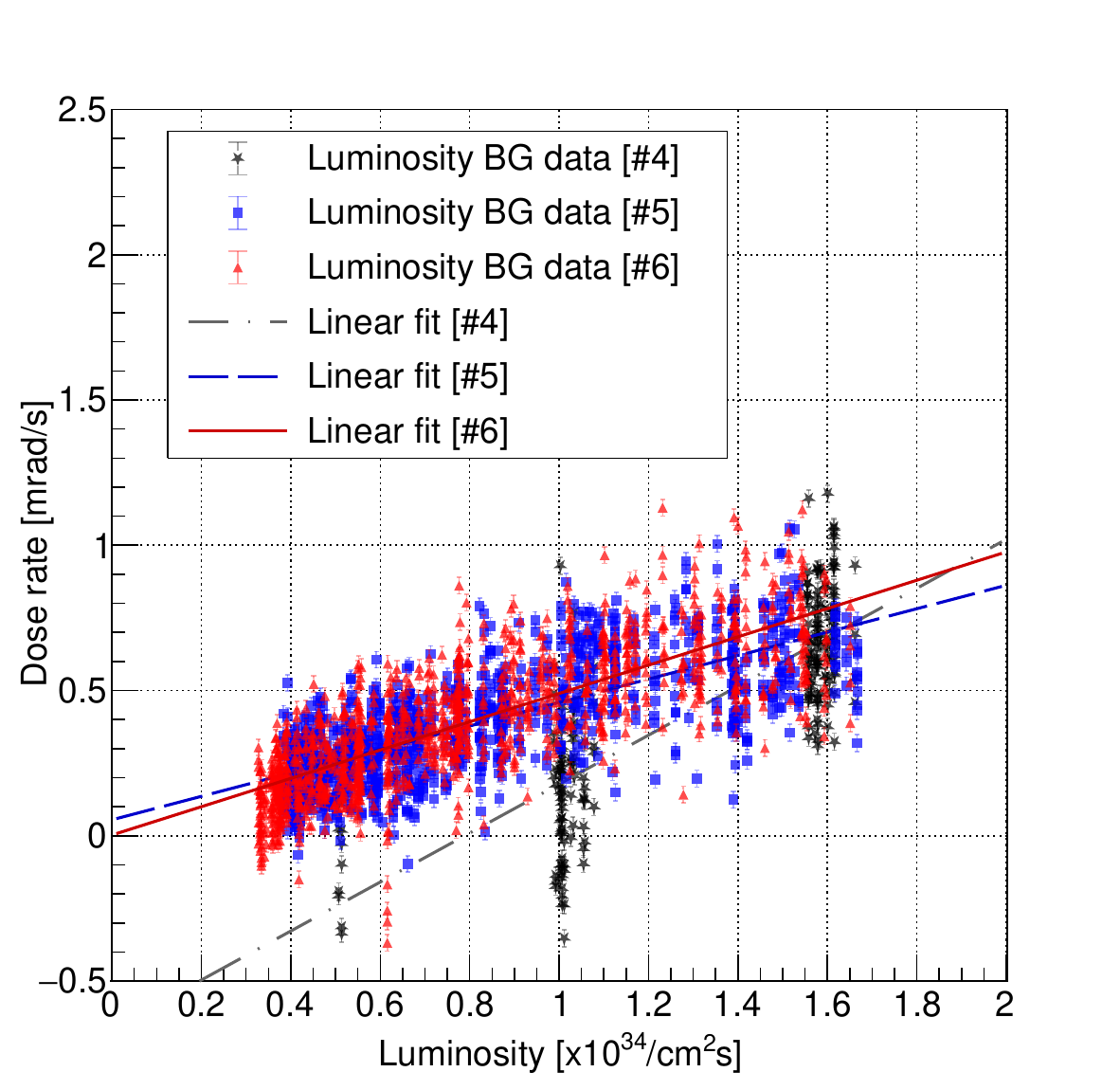}
\caption{\label{secC:fig2}Luminosity component of the measured BP-FW-325 diamond detector dose rate versus instantaneous luminosity from the May~9, 2020 study.}

\end{figure}

Figure~\ref{secC:fig2} shows the luminosity background versus the collision luminosity, measured by the ECL as explained in Section~\ref{sec:Introduction}, for the top-up injection period (\#4, solid black stars) and the two beam decays (\#5, solid blue squares, and \#6, solid red triangles). As expected, these three distributions illustrate a clear linear dependency between the luminosity background ($\mathcal{O}_\mathrm{lumi}$) and the instantaneous luminosity ($\mathcal{L}$). We fit the estimated luminosity background versus luminosity with a first-order polynomial, as shown in Fig.~\ref{secC:fig2}. In the absence of any residual systematic effects, we would expect all three fits to go through the origin and to have very similar slopes. For the particular Diamond detector shown in Fig.~\ref{secC:fig2}, this is the case for the fits to data sets \#5 and \#6. The fit to data set \#4, however, has a different slope and a negative intercept with the vertical axis, which would correspond to negative luminosity background and is unphysical. We speculate that for this detector, data set \#4 is biased by a residual contribution of the injection background that leaks into our estimated non-injection background. In addition, our analysis implicitly assumes that the non-luminosity background sensitivities are the same during single-beam and collision modes of the accelerator. If this assumption does not hold, offsets such as observed in data set \#4 are also possible. To account for these uncertainties, the final luminosity background extrapolation for all detectors discussed in the text below generally uses the average slope of three linear fits analogous to those shown in Fig.~\ref{secC:fig2}, with the caveat that fits with negative slope are discarded. For each detector, we assign a systematic uncertainty equal to the average of the three (or fewer, if some of the fits are discarded) intercepts with the vertical axis.

\subsection{Injection background}

SuperKEKB requires continuous injection of particles to keep the beam currents constant and luminosity high. Beam losses in the IR can increase for a short period of time, typically $\mathcal{O}(\SI{10}{ms})$, after injection, which can be detrimental to both detector operations and reconstruction performance. In order to avoid DAQ saturation, a L1-trigger veto rejects triggers that occur close to the time when a newly injected bunch passes the IP. Therefore, in most detectors only the part of the injection background that is outside the L1-trigger veto is seen as an excess over the storage (non-injection) background. However, vetoed events will still contribute to the dose rate seen by detectors, and hence must be included in dose rate estimates. Prediction of the injection background via simulation is a very challenging task, as it depends on a broad spectrum of machine parameters, all the way from the particle gun and LINAC to the injector and stored beam.

Below, we compare two methods under development to estimate the SuperKEKB injection background in Belle~II experimental data.

\subsubsection{Background remnant}

One straightforward approach to estimating the injection background ($\mathcal{O}_\mathrm{inj.}$) is to use the heuristic fit results described above. We subtract the estimated storage background ($\mathcal{O}_\mathrm{est.}$) from measured data ($\mathcal{O}_\mathrm{meas.}$) during a top-up injection period of \SI{5}{min} before each beam decay. 

\begin{figure}[htbp]
\centering
\includegraphics[width=\linewidth]{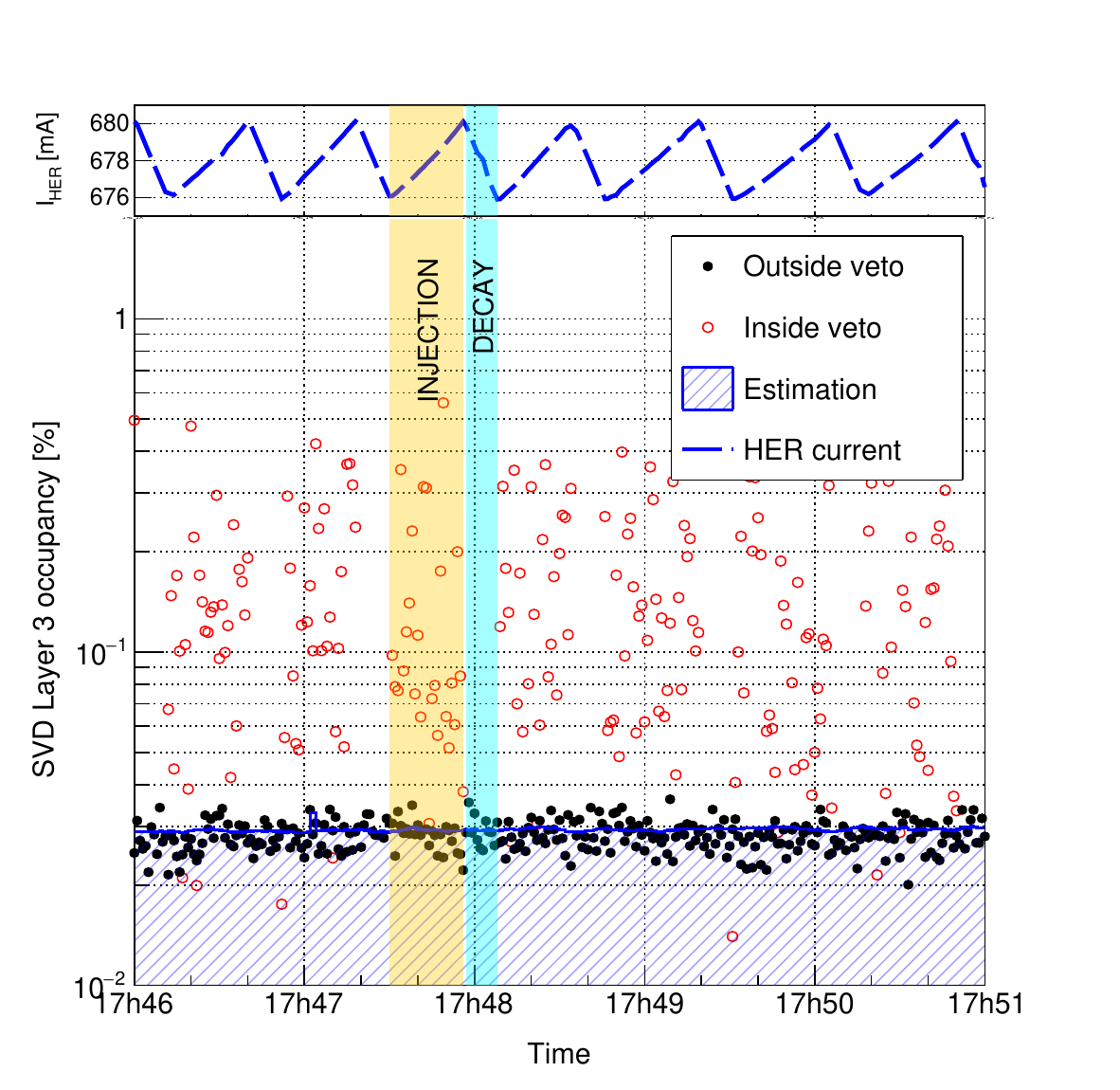}
\caption{\label{secC:fig4}Top: measured HER beam current during top-up injection for June~16, 2021 background studies; bottom: measured occupancy for the inner-most SVD layer.}

\end{figure}

Figure~\ref{secC:fig4} illustrates the measured background for the SVD L3 during the HER single-beam top-up injection. The upper part of the figure shows the HER beam current with 1174 bunches of electrons. The bottom part of the figure contains two data sets of the measured mean occupancy with a timestamp of \SI{1}{Hz} for outside (black, solid circles) and inside (red, open circles) the injection veto window. The blue, hatched area represents the estimated HER storage background extrapolated by using heuristic fit results ($\mathcal{O}_\mathrm{single}^\mathrm{HER}$). Seven beam injection periods occur in this figure, where the 1-bunch injection repetition rate is \SI{12.5}{Hz}. One of the injection periods is highlighted by a vertical orange band. The subsequent beam decay period is highlighted in cyan. The frequency of injection periods depends on the beam lifetime and the maximum acceptable beam current drop, typically set at 1\% of the operational current. The figure is a good illustration of the injection trigger veto performance. The trigger system vetoes high beam losses for about \SI{10}{ms} right after the beam injection inside the veto window to ensure stable DAQ operation. When the injection is stopped, the beam current decays (vertical cyan band in the figure), and the observed background is presumably due to the storage beam circulating in the ring. 

To estimate the full radiation dose (and hence the potential for radiation damage of electronics) on Belle~II sub-detectors, the contribution from injection background, including the component \textit{hidden} by the L1-trigger injection veto, must be included. Data inside the trigger veto window is affected by the DAQ dead time fraction due to the veto, $F_\mathrm{DT} \sim 3-6\%$. Furthermore, we only inject the beam some fraction of the time (see Fig.~\ref{secC:fig4}), $F_\mathrm{ID} \sim 50-70\%$, which is defined as the ratio of the injection duration to the sum of the injection duration and decay duration. Both $F_\mathrm{DT}$ and $F_\mathrm{ID}$ must be accounted for when normalizing the estimated injection background.

We define the relative injection background as
\begin{equation}
    \mathcal{\widetilde{O}}_\mathrm{inj.} = \mathcal{O}_\mathrm{inj.}/\mathcal{O}_\mathrm{est.} = (\mathcal{O}_\mathrm{meas.} - \mathcal{O}_\mathrm{est.})/\mathcal{O}_\mathrm{est.}.
    \label{secC:eq7}
\end{equation}
Since the injection background is seen only during a short period when a fresh beam is injected into the main ring, each data point in Fig.~\ref{secC:fig4} is then normalized by $F_\mathrm{DT}$ and $F_\mathrm{ID}$: 
\begin{equation}
    \mathcal{\widetilde{O}}_\mathrm{inj.}^\mathrm{norm.,in} = \mathcal{\widetilde{O}}_\mathrm{inj.}^\mathrm{in} \times F_\mathrm{DT} \times F_\mathrm{ID},
    \label{secC:eq11}
\end{equation}
\begin{equation}
    \mathcal{\widetilde{O}}_\mathrm{inj.}^\mathrm{norm.,out} = \mathcal{\widetilde{O}}_\mathrm{inj.}^\mathrm{out} \times (1 - F_\mathrm{DT}) \times F_\mathrm{ID},
    \label{secC:eq12}
\end{equation}
 where $\mathcal{\widetilde{O}}^\mathrm{norm.}_\mathrm{inj.}$ is the normalized injection fraction. 
 
\begin{figure}[!ht]
\centering
\includegraphics[width=\linewidth]{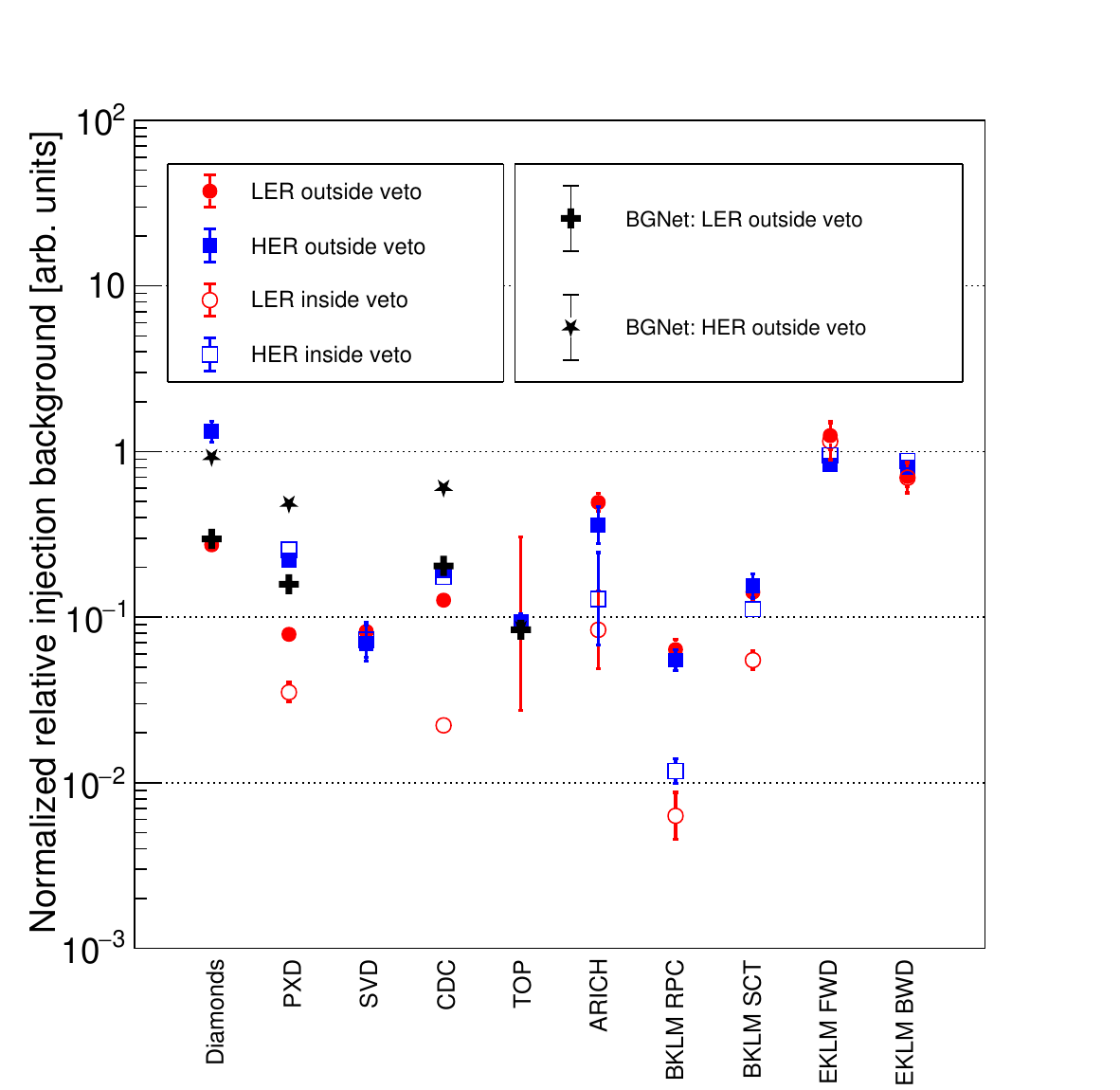}
\caption{\label{secC:fig5}The normalized relative injection background during the June~2021 study. 
BKLM~RPC stands for the barrel KLM layers made of RPCs, BKLM~SCT represents barrel KLM scintillator layers, and EKLM~FWD and BWD show the relative injection background for the KLM endcaps on the forward and backward sides, respectively, made of scintillator layers.}

\end{figure}

 Figure~\ref{secC:fig5} shows the Belle~II normalized relative injection background for the June~2021 study, where $\mathcal{\widetilde{O}}^\mathrm{norm.}_\mathrm{inj.}$ varies within one order of magnitude for outside ($\mathcal{\widetilde{O}}_\mathrm{inj.}^\mathrm{norm.,out}$, solid markers) and inside ($\mathcal{\widetilde{O}}_\mathrm{inj.}^\mathrm{norm.,in}$, open markers) the injection veto data samples. The sampled beam background data with a timestamp of \SI{100}{ms} and \SI{1}{s} for the diamond and TOP detectors, respectively, are collected without the L1-trigger and injection veto. Each data point in Fig.~\ref{secC:fig5} is normalized by the DAQ dead time fraction during the injection and decay periods following Eqs.~\eqref{secC:eq11} and \eqref{secC:eq12}, respectively. Error bars illustrate the total uncertainty, including statistical and systematic errors, where the latter is calculated as a geometric standard error over all layers in a given sub-detector.

Based on the measured total and estimated storage backgrounds, this method allows us to estimate the injection background during top-up injection in one of the rings. Following this approach, we can evaluate the radiation damage in Belle~II sub-detectors by integrating the storage and injection background doses, taking the injection trigger veto impact into account, and properly normalising the injection background fraction. However, the main limitation of this method is that the fraction of the injection background does not stay constant for a long time during machine operation due to continuous machine tuning and different beam (bunch) currents and luminosity. Therefore, the results of this injection background estimation can be extrapolated outside the dedicated background study runs only with certain assumptions regarding the ratio between injection and storage background components measured by the detector.

\subsubsection{Neural network}

BGNet~\cite{Schwenker2023} is an artificial neural network for predicting the background rate of Belle~II sub-detectors. The network learns to map SuperKEKB collider variables to background hit rates caused by different beam background sources seen by Belle~II. One major motivation is to accurately extract background hit rates from top-up injections, understand their dependence on collider conditions, and mitigate their impact on data taking.  Feature attribution algorithms~\cite{Sundararajan2017,Erion2019} are applied to identify the most predictive input variables. 

BGNet consists of neural network-based models for the most relevant background sources as the physical origin for the loss of beam particles near the interaction region of Belle~II as follows: i) the beam-gas storage background in the LER and HER, ii) the Touschek storage background in the LER and HER, iii) the luminosity background, iv) the LER and HER top-up injection background, and v) detector pedestals. The models for beam-gas and Touschek contributions to the hit rate follow Eq.~\eqref{secC:eq1} and Eq.~\eqref{secC:eq2} but replace the coefficients $B$ and $T$ by fully connected feed-forward artificial neural networks, respectively.  The injection background hit rate network (separately for the HER/LER) is a fully connected feed-forward network multiplied with an injection gate status variable. The injection gate status is open (variable value of 1) whenever top-up injections into a ring take place, otherwise, it is closed (value of 0). The collision and pedestal-related background components are represented by the weight and bias of a linear neuron with the measured luminosity as its only input variable.       

BGNet is trained on archived \SI{1}{Hz} time series of process variables (PVs) provided by the EPICS-based slow-control system of Belle~II. The training target is the observed total background hit rate of a Belle~II sub-detector. The input tensors for HER/LER injection and storage background networks are selected based on expert knowledge, and the result of feature attribution methods is used to rank the importance of variables. During training, BGNet optimizes the weights and biases of its sub-networks to minimize the mean absolute error between the measured hit rate and the sum over all predicted background components. The data are split into training and validation sets. All input variables and the measured hit rate are scaled by subtracting the median and scaling by the percentile range between the 90th and 10th percentile. 

\begin{figure}[htbp]
\centering
\includegraphics[width=\linewidth]{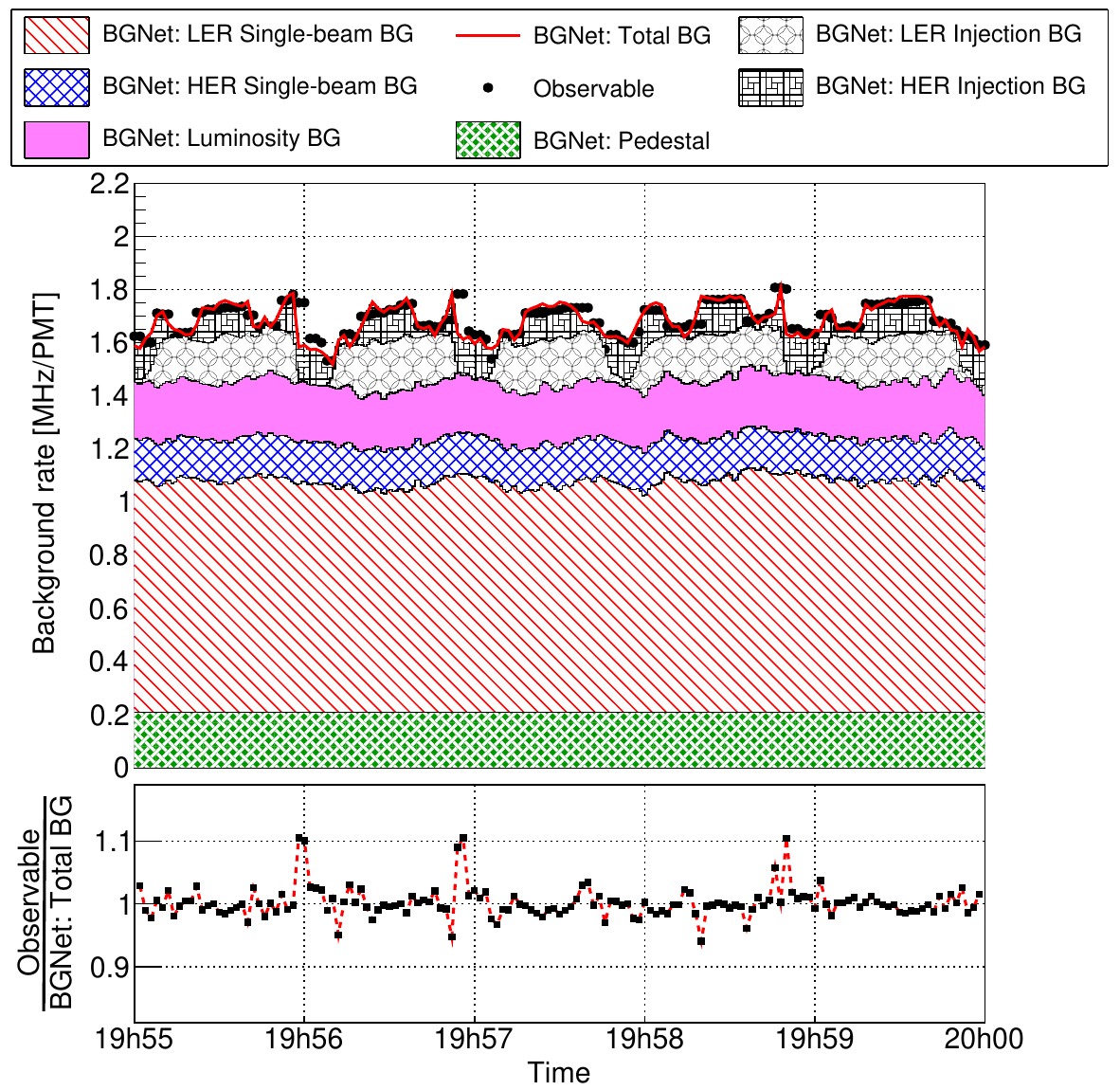}
\caption{\label{secC:fig8}Components of the TOP detector background predicted by BGNet for the June~16, 2021 background study. Top: stacked histograms of predicted background components displayed on top of the observable; bottom: the ratio between the observable and predicted total background.}

\end{figure}

Figure~\ref{secC:fig5} contains BGNet estimation results for the relative injection background outside the veto. The neural network and heuristic fit results demonstrate an acceptable agreement for the outside veto data. However, there is a noticeable disagreement for some sub-detectors, e.g. for the PXD and CDC, since BGNet uses the online archived data, which may contain noisy electronics channels that are masked for the heuristic fit offline. 

Storage backgrounds (single-beam and luminosity) are learned mostly from beam decay data during single-beam and collision operation of the collider. During physics runs, the injection backgrounds show a typical temporal pattern following the injection gate status in the HER and LER since the top-up injections regularly paused and resumed to keep the beam currents constant, as shown in Fig.~\ref{secC:fig8}, which corresponds to the following beam parameters: $I^\mathrm{LER/HER}=\SI[parse-numbers = false, number-math-rm = \ensuremath]{740/650}{mA}$, $n_\mathrm{b}=1174$, and $\mathcal{L} = \SI{2.6e34}{cm^{-2}s^{-1}}$. The injections into the HER and LER are asynchronous. The contribution of HER and LER injections can be disentangled even during physics runs by looking at the beam gate status variables.

The BGNet was tested on recorded data during Belle~II operation in 2021 and 2022. After training, the model learned a physically sensible and accurate decomposition of the detector observables into components for different background sources. In addition, feature attribution algorithms have been applied to the sub-models in BGNet to understand which inputs the sub-models find most valuable for making predictions.
The method can provide valuable clues to understand the backgrounds in Belle~II better. We are working on further developing the neural network to make it a helpful tool used by SuperKEKB operators for crucial machine parameter tuning, to mitigate backgrounds, or to improve collider performance.

\section{\label{sec:SummaryOfCurrentBackgroundLevels}Summary of the measured background composition}

This section summarizes the background status in Belle~II as of June~2021, reporting on our current understanding of beam-induced backgrounds. At that time, the detector was running with stable machine operation with well-controlled and understood beam backgrounds, in contrast to 2022 operation with frequent sudden beam losses and damaged collimators. We also compare background measurements against dedicated simulations.

\subsection{Measured backgrounds}

Table~\ref{secD:tab1} shows detector limits. The TOP limit before LS1 is related to the replacement of TOP conventional PMTs planned for LS1. At the same time, the limit after LS1 is associated with the replacement of ALD PMTs in LS2 and the longevity of life-extended ALD PMTs. Moreover, the upper background rate limit quoted for the Diamond read-out electronics can be increased by selecting a lower signal amplification. The KLM detector limit corresponds to the muon reconstruction efficiency drop of about 10\%.

The estimated future background in Table~\ref{secD:tab1} is the main goal of this article, and obtaining this requires knowledge of the detailed background composition and good Data/MC agreement. These topics will be detailed in what follows.

\begin{table*}[htbp]
\centering
    \caption{\label{secD:tab1}Background rate limits for different Belle~II detector sub-systems. The third column shows the total measured background rate in June~2021 at $\mathcal{L} = \SI{2.6e34}{cm^{-2}.s^{-1}}$ excluding the pedestal rate. The fifth column shows the total estimated background rate before LS2 at $\mathcal{L} = \SI{2.8e35}{cm^{-2}.s^{-1}}$. The TOP luminosity background is assumed to be \SI{0.925}{MHz/PMT} per $10^{35}\rm~cm^{-2}s^{-1}$.}
    \begin{tabular}{lcccccc}
    \hline\hline
    Detector & \multicolumn{2}{c}{BG rate limit} & \multicolumn{2}{c}{Current (June 2021)} & \multicolumn{2}{c}{Estimated (Before LS2)} \\
    & & & Background & Safety factor & Background & Safety factor \\
    \hline
    Diamonds & \multicolumn{2}{c}{\SIrange{1}{2}{rad/s}} & $<\SI{132}{mrad/s}$ & $>17$ & $<\SI{311}{mrad/s}$ & $>7.2$\\
    PXD & \multicolumn{2}{c}{3\%} & 0.1\% & 30.1 & 0.4\% & 6.9 (L1)\\
    SVD L3, L4, L5, L6 & \multicolumn{2}{c}{4.7\%, 2.4\%, 1.8\%, 1.2\%} & $<0.22\%$ & 21.5 & 1.0\% & 4.7 (L3)\\
    CDC & \multicolumn{2}{c}{\SI{150}{kHz/wire}} & \SI{22.3}{kHz/wire} & 6.7 & \SI{79}{kHz/wire} & 1.9\\
    ARICH & \multicolumn{2}{c}{\SI{10}{MHz/HAPD}} & \SI{0.5}{MHz/HAPD} & 21.7 & \SI{1.4}{MHz/HAPD} & 7.3\\
    Barrel KLM L3 & \multicolumn{2}{c}{\SI{50}{MHz}} & \SI{4}{MHz} & 12.1 & \SI{12}{MHz} & 4.1\\
    & \multicolumn{2}{c}{non-luminosity BG} &\\\cline{2-3}
    & before LS1 & after LS1 &\\\cline{2-3}
    TOP & \SI{3}{MHz/PMT} & \SI{5}{MHz/PMT} & \SI{1.8}{MHz/PMT} & 1.8 & \SI{5.0}{MHz/PMT} & 1.5\\
    & \multicolumn{2}{c}{+ luminosity BG}& & \\
    \hline\hline
    \end{tabular}
\end{table*}

\begin{figure*}[htbp]
\centering
\includegraphics[width=\linewidth]{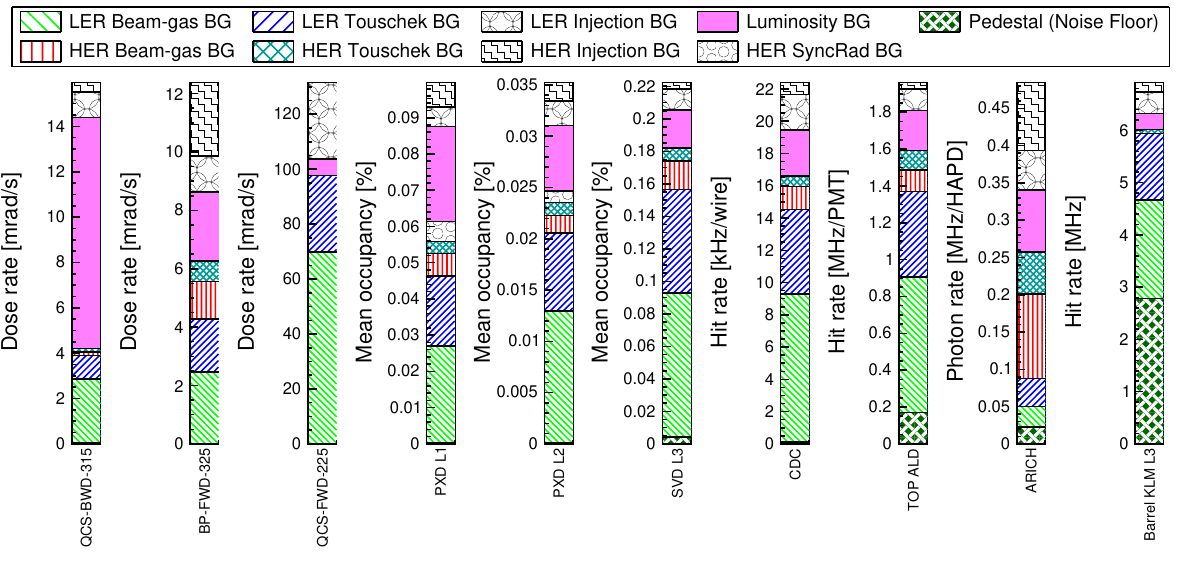}
\caption{\label{secD:fig1}Measured Belle~II background composition on June~16, 2021. Each column is a stacked histogram. 
QCS-BWD-315, BP-FWD-325 and QCS-FWD-225 indicate backward QCS, beam pipe and forward QCS Diamond detectors, respectively, with the higher dose rate. Barrel~KLM~L3 corresponds to the inner-most RPC layer in the barrel region of the KLM detector. TOP~ALD shows the averaged background over ALD-type MCP-PMTs, slots from 3 to 9.
}

\end{figure*}

Figure~\ref{secD:fig1} shows the measured background rate and composition (i.e. decomposed by the most significant beam loss sources) for each Belle~II sub-system separately. The data used are from the luminosity background study on June~16, 2021 at the following beam condition: $I^\mathrm{LER/HER}=\SI[parse-numbers = false, number-math-rm = \ensuremath]{732.6/647.2}{mA}$, $n_\mathrm{b}=1174$, $\sigma_\mathrm{x}^\mathrm{LER/HER}=\SI[parse-numbers = false, number-math-rm = \ensuremath]{184.6/151.0}{\upmu m}$, $\sigma_\mathrm{y}^\mathrm{LER/HER}=\SI[parse-numbers = false, number-math-rm = \ensuremath]{60.7/36.2}{\upmu m}$, $\sigma_\mathrm{z}^\mathrm{LER/HER}=\SI[parse-numbers = false, number-math-rm = \ensuremath]{6.5/6.8}{mm}$, $P_\mathrm{eff.}^\mathrm{LER/HER}=\SI[parse-numbers = false, number-math-rm = \ensuremath]{88.7/24.3}{nPa}$, and $\mathcal{L} = \SI{2.6e34}{cm^{-2}.s^{-1}}$. Beam-gas, Touschek, luminosity, and PXD SR backgrounds are obtained using the heuristic fit methodology described earlier. The total injection background ($\mathcal{O}_\mathrm{inj.}$) corresponds to the inside ($\mathcal{\widetilde{O}}_\mathrm{inj.}^\mathrm{norm.,in}$, Eq.\ref{secC:eq11}) and outside ($\mathcal{\widetilde{O}}_\mathrm{inj.}^\mathrm{norm.,out}$, Eq.\ref{secC:eq12}) the veto injection background normalized by the DAQ dead time and injection duration fractions during top-up injection and beam decay:

\begin{equation}
    \mathcal{O}_\mathrm{inj.} = ( \mathcal{\widetilde{O}}_\mathrm{inj.}^\mathrm{norm.,in}+ \mathcal{\widetilde{O}}_\mathrm{inj.}^\mathrm{norm.,out})\times \mathcal{O}_\mathrm{single},
    \label{secD:eq1}
\end{equation}
where $\mathcal{O}_\mathrm{single}$ is the estimated single-beam background.

The overall background level for all sub-systems is well below the detector limits listed in Table~\ref{secD:tab1}.  The dominant backgrounds are due to LER beam-gas, LER Touschek and luminosity beam losses. HER and injection backgrounds are much lower, at the level of 10\%, except for the ARICH, which is more sensitive to FWD-directed beam losses from the HER beam. The reported rates are affected by the so-called event-of-doom buster (EoDB), introduced in 2020. The EoDB removes events with more than $\SI{6000}{hits}$ in the CDC or more than $\SI{70000}{digits}$ in the SVD, introducing a systematic bias of about 20\% to the measured total background rate during the injection.

At the present level, the SR background is of no concern in terms of occupancy for the inner-most layers of the vertex detector. However, its potential increase at higher beam currents or at different beam orbits tuned to increase the luminosity may cause inhomogeneities in the irradiation of the PXD modules, which is difficult to compensate by simply adjusting the operation voltages of the affected modules.

\begin{table}[htbp]
\centering
    \caption{\label{secD:tab3}The measured fast neutron background by TPCs in the accelerator tunnel at $\mathcal{L} = \SI{2.6e34}{cm^{-2}.s^{-1}}$.}
    \begin{tabular}{cccc}
    \hline\hline
    Background & Accelerator & Fluence per smy \\
    type & tunnel & [$\times 10^{8}\rm~n_\mathrm{eq}/cm^2$]\\
    \hline
    Single-beam & BWD/FWD & 6/90 \\
    Luminosity & BWD/FWD & 40/4 \\
    \hline\hline
    \end{tabular}
\end{table}

\begin{table}[htbp]
\centering
    \caption{\label{secD:tab4}The measured thermal neutron background by \ce{^3He}~tubes in the accelerator tunnel at $\mathcal{L} = \SI{2.6e34}{cm^{-2}.s^{-1}}$.}
    \begin{tabular}{cccc}
    \hline\hline
    Background & Accelerator & Flux \\
    type & tunnel & [$\times 10^{2}\rm~n/(cm^2 s)$]\\
    \hline
    Single-beam & BWD/FWD & 1/30 \\
    Luminosity & BWD/FWD & 20/4 \\
    \hline\hline
    \end{tabular}
\end{table}

The neutron background is not considered explicitly in the study reported here. However, the neutron background in the SuperKEKB tunnel near Belle~II has been studied separately, using direction and energy-sensitive gas TPC detectors to image neutron recoils~\cite{Schueler2021}. Those results, converted into estimated 1-MeV neutron equivalent fluences per Snowmass year, are listed in Table~\ref{secD:tab3}. In addition, Table~\ref{secD:tab4} reports on the current thermal neutron fluxes measured by the \ce{^3He}~tube neutron counting system, which were not previously published.

TPC and \ce{^3He}~tube results show that LER single-beam backgrounds are the dominant background sources in the FWD tunnel, which could be explained by high beam losses at the nearest, tightly closed collimator $\sim\SI{16}{m}$ from the IP, see Fig.~\ref{secB:fig4}. On the other hand, the BWD tunnel neutrons are predominantly due to luminosity background ``hotspots'', which are expected on either side of the Belle~II detector~\cite{Schueler2021}.

Moreover, the neutrons from the electromagnetic showers, originating from both the IR and accelerator tunnels, might be the reason for SEUs of FPGA electronics boards seen during the beam operation. Our simulation and dedicated machine studies show that beam losses at the collimators nearest to the detector, and thus single-beam neutrons, can be suppressed by aperture adjustment of  distant upstream collimators in each ring. However, we can only mitigate the luminosity neutron background by installing additional shielding around the detector. We are currently working on further neutron background studies, dedicated countermeasures, and possible detector upgrades, which will be discussed in forthcoming publications.

\subsection{Simulation accuracy}

\begin{figure*}[htbp]
\centering
\subfloat[\label{secD:fig2a}LER single-beam background.]{\includegraphics[width=0.33\linewidth]{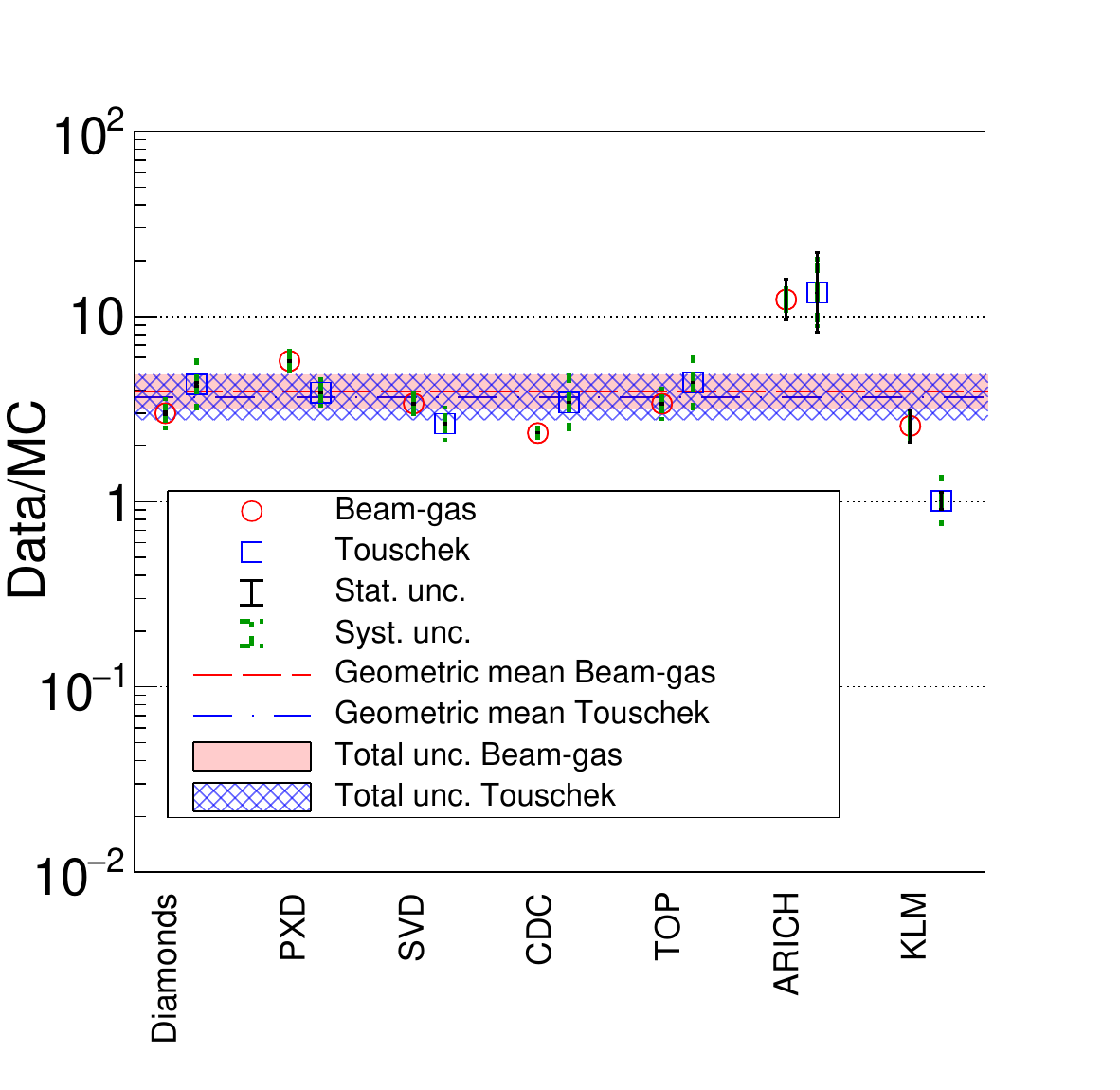}}
\hfill
\subfloat[\label{secD:fig2b}HER single-beam background.]{\includegraphics[width=0.33\linewidth]{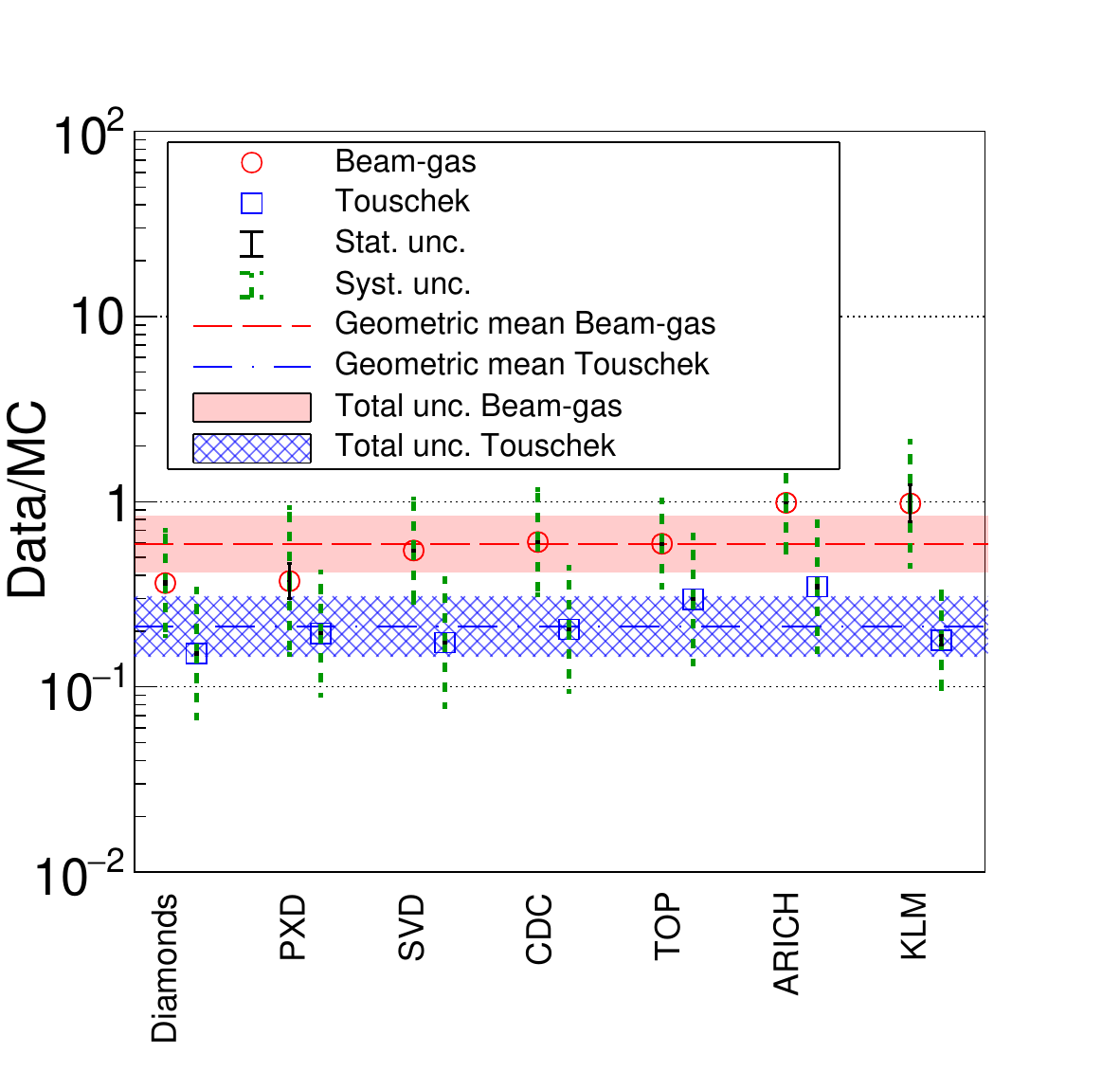}}
\hfill
\subfloat[\label{secD:fig2c}Luminosity background.]{\includegraphics[width=0.33\linewidth]{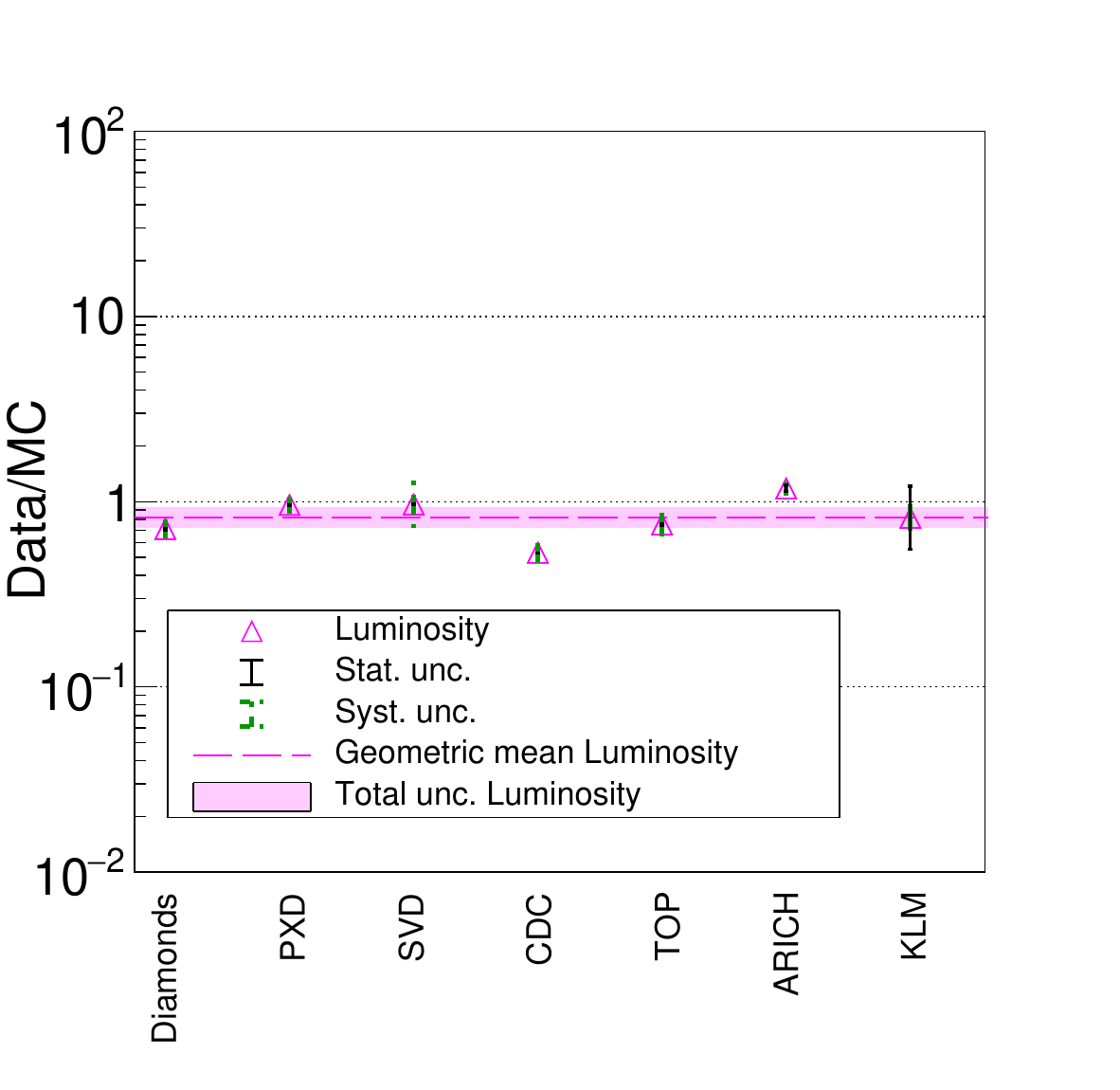}}
\caption{\label{secD:fig2}Belle~II detector-level Data/MC ratios in Belle~II over 2020 and 2021 dedicated background studies.
}

\end{figure*}

To probe the accuracy of the Belle~II background simulation and our current understanding of the major beam loss processes in SuperKEKB, we calculate the Data/MC ratio for the four beam background studies performed in 2020 and 2021 (see Section~\ref{subsec:DedicatedBgStudy}).
A dedicated set of Monte-Carlo simulations based on SAD and Geant4 is prepared for each study using the procedure discussed in Section~\ref{sec:BackgroundSimulation}. Each component of the measured background is then scaled to the simulated beam parameters using the heuristic fit results so that measured and simulated rates can be compared for identical beam parameters. Figure~\ref{secD:fig2} shows a summary of the findings, Belle~II detector-level Data/MC ratios, where each value is calculated as a geometric mean over i) the relevant sub-detector's layers, modules, sensors or segments as discussed in Section~\ref{sec:MeasuredBackgroundComposition}, and over ii) the four background studies. The statistical uncertainties originate from the heuristic fit parameter errors, while the systematic uncertainties are defined as variations of the individual ratio around the mean value and calculated as a standard error of the geometric mean~\cite{Norris1940,Alf1979}. The measured and simulated data are compared at arbitrary beam parameters: $I^\mathrm{LER/HER}=\SI[parse-numbers = false, number-math-rm = \ensuremath]{1.2/1.0}{A}$, $n_\mathrm{b}=1576$, $\mathcal{L} = \SI{8e35}{cm^{-2}.s^{-1}}$. The average gas pressure is estimated based on reported parameters in Table~\ref{secC:tab2}. The combined ratios over all Belle~II sub-systems for single-beam and luminosity backgrounds are summarized in Table~\ref{secD:tab2}.

\begin{table}[htbp]
\centering
    \caption{\label{secD:tab2}Combined Belle~II Data/MC ratios over 2020-2021 collected data.}
    \begin{tabular}{ccc}
    \hline\hline
    Background & LER & HER \\
    \hline
    Beam-Gas & ${3.94}_{-0.74}^{+0.92}$ & ${0.59}_{-0.18}^{+0.25}$ \\
    Touschek & ${3.67}_{-0.92}^{+1.22}$ & ${0.21}_{-0.07}^{+0.10}$ \\
    Luminosity & \multicolumn{2}{c}{${0.82}_{-0.10}^{+0.11}$} \\
    \hline\hline
    \end{tabular}
\end{table}

As reported in Refs.~\cite{LEWIS201969} and \cite{Liptak2021}, agreements between first measurements and optimistic background simulation in 2016 and 2018 were poor, and Data/MC ratios strongly diverged from the unity by several orders of magnitude. Therefore, during the early Phase~3 discussed in this paper, we invested a lot of effort in improving the beam-induced background simulation for a better understanding of beam loss mechanisms in the machine. The main key improvements compared to Phase~1 and Phase~2, leading to the substantial measurement and simulation agreement, are i) the realistic collimator profile implementation in SAD, ii) particle interaction with collimator materials (tip-scattering), iii) beam-gas losses re-weighting using the measured vacuum pressure distribution around the SuperKEKB rings, iv) accurate translation of lost particle coordinates from SAD to Geant4, and v) the improved Geant4 model of the machine and detector components and the accelerator tunnel.

\section{\label{sec:Extrapolations}Extrapolations}

This section estimates the expected detector background at higher luminosity based on a dedicated set of Monte-Carlo simulations. These simulations help us study machine and detector upgrades needed to achieve the planned machine performance. Below, we review our methodology for extrapolating the beam backgrounds to a luminosity of \SI{2.8e35}{cm^{-2}.s^{-1}}, which is expected to be achieved by January~2027, before the start of LS2.

\begin{table}[htbp]
\centering
    \caption{\label{secE:tab2}Predicted SuperKEKB parameters, expected to be achieved by the specified date. $\beta^\mathrm{*}$, $\mathcal{L}$, $I$, $BD_\mathrm{int.}$, $n_\mathrm{b}$, $\varepsilon$, $\sigma_\mathrm{z}$ and $CW$ stand for the betatron function at the IP, luminosity, beam current, integrated beam dose, number of bunches, equilibrium beam emittance, bunch length and Crab-Waist sextupoles, respectively.}
    \begin{tabular}{lccccc}
    \hline\hline
    Setup & Before LS2 & Target\\
    \hline
    Date & Jan~2027 & Jan~2031 \\
    $\beta^\mathrm{*}_\mathrm{y}$(LER/HER)~[mm]& 0.6/0.6 & 0.27/0.3 \\
    $\beta^\mathrm{*}_\mathrm{x}$(LER/HER)~[mm]& 60/60 & 32/25 \\
    $\mathcal{L}$~[$\times 10^{35}\rm~cm^{-2}s^{-1}$]& 2.8 & 6.0 \\
    $I$(LER/HER)~[A] & 2.52/1.82 & 2.80/2.00 \\
    $BD_\mathrm{int.}$~[kAh] & 45 & 93 \\
    $n_\mathrm{b}$~[bunches] & 1576 & 1761 \\
    $\varepsilon_\mathrm{x}$(LER/HER)~[nm] & 4.6/4.5 & 3.3/4.6 \\
    $\varepsilon_\mathrm{y}/\varepsilon_\mathrm{x}$(LER/HER)~[\%] & 1/1 & 0.27/0.28 \\
    $\sigma_\mathrm{z}$(LER/HER)~[mm] & 8.27/7.60 & 8.25/7.58 \\
    $CW$ & OFF & OFF \\
    \hline\hline
    \end{tabular}
\end{table}

To collect an integrated luminosity of \SI{50}{ab^{-1}} by the 2030s, our target instantaneous luminosity at $\beta^{*}_\mathrm{y}=\SI{0.3}{mm}$ is \SI{6e35}{cm^{-2}.s^{-1}}. Table~\ref{secE:tab2} lists predicted future beam parameters based on the most recent SuperKEKB plan for ramping up the machine~\cite{Suetsugu2021prv}. Unfortunately, with the machine lattice considered in the original machine design without the Crab-Waist scheme~\cite{Ohnishi2013}, the target beam currents will be difficult or even impossible to reach because of the short beam lifetime ($<\SI{10}{min}$) due to the narrow dynamic aperture~\cite{Morita2015}. Moreover, our preliminary estimates show that it may be challenging to safely run the experiment at the target beam parameters due to the low TMCI bunch current threshold for narrow collimator apertures. Thus, we might be forced to open some collimators, which could increase the IR background above the detector limits. In Ref.~\cite{Natochii2022}, we have proposed a few possible solutions to partially cure beam instabilities and resolve the specific luminosity and dynamic aperture degradation, where the latter affects beam lifetime, as mentioned above. Nevertheless, the upshot is that the target machine lattice and beam parameters are still too uncertain to make an accurate background prediction for the target luminosity. Therefore here, we focus on estimating backgrounds for intermediate beam parameters, which are feasible to achieve before LS2. In our simulations, the Crab-Waist scheme is not used, resulting in conservative background estimates. According to preliminary, separate SAD-only simulations, the Crab-Waist scheme at $\beta^{*}_\mathrm{y} = \SI{0.6}{mm}$, is expected to lower Belle~II beam backgrounds by at least a factor of three, if simulation-optimized collimator settings can be achieved experimentally. 

To project the beam-gas background forward in time, we start by extrapolating the beam pipe pressure measurements performed in 2021. Next  the collimator system configuration is optimized in simulation to reduce single-beam backgrounds in the IR while maintaining an acceptable beam lifetime. Finally, we estimate all simulated background components in each sub-detector, which are then scaled by corresponding Data/MC ratios discussed above to estimate the expected background level. This results in limits on beam pipe vacuum pressure, injection quality, and collimation, which must be achieved to keep the background in Belle~II sub-detectors below their rate limits.

\subsection{Gas pressure}

\begin{figure*}[htbp]
\centering
\subfloat[\label{secE:fig1a}LER.]{\includegraphics[width=0.48\linewidth]{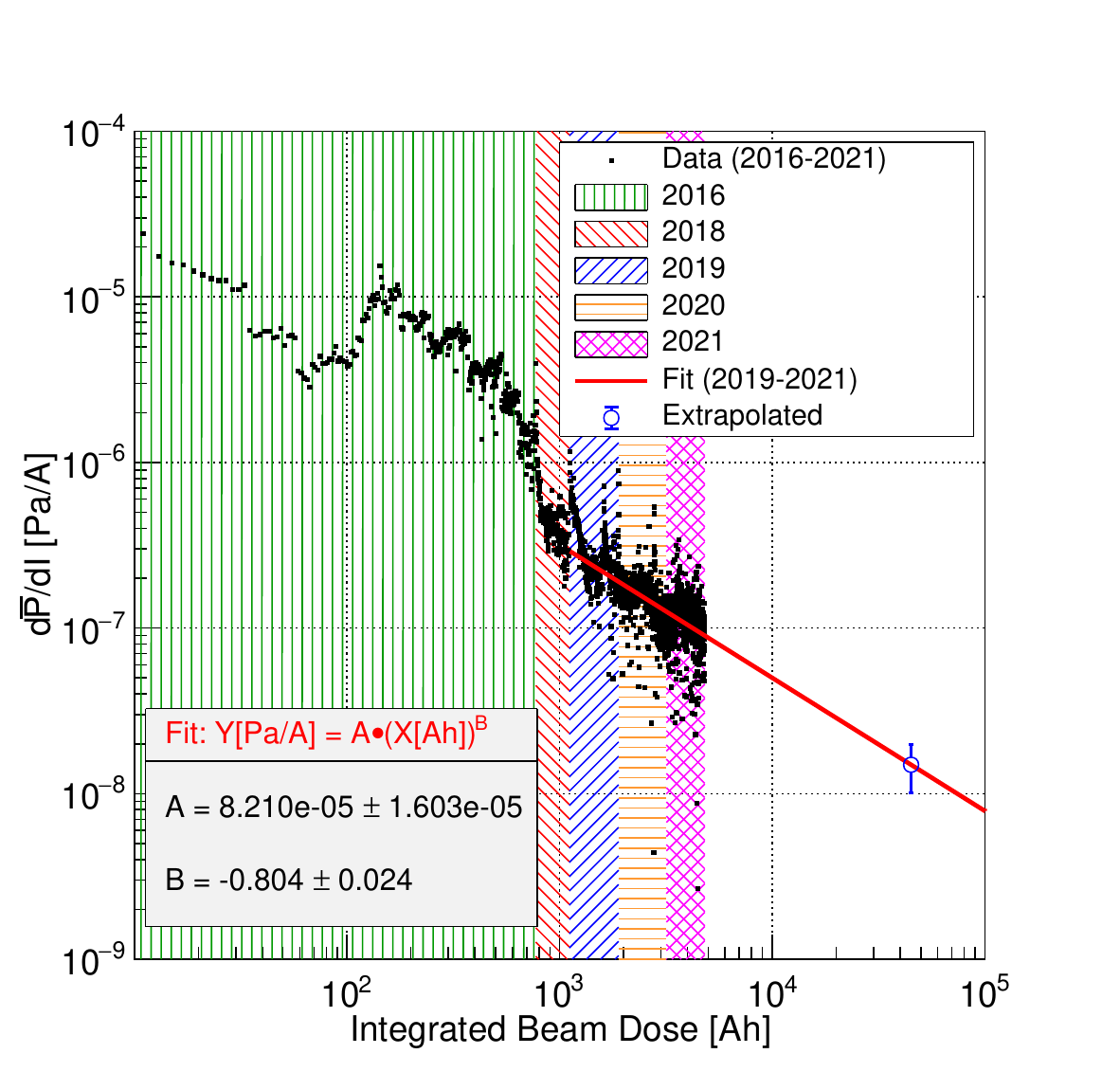}}
\hfill
\subfloat[\label{secE:fig1b}HER.]{\includegraphics[width=0.48\linewidth]{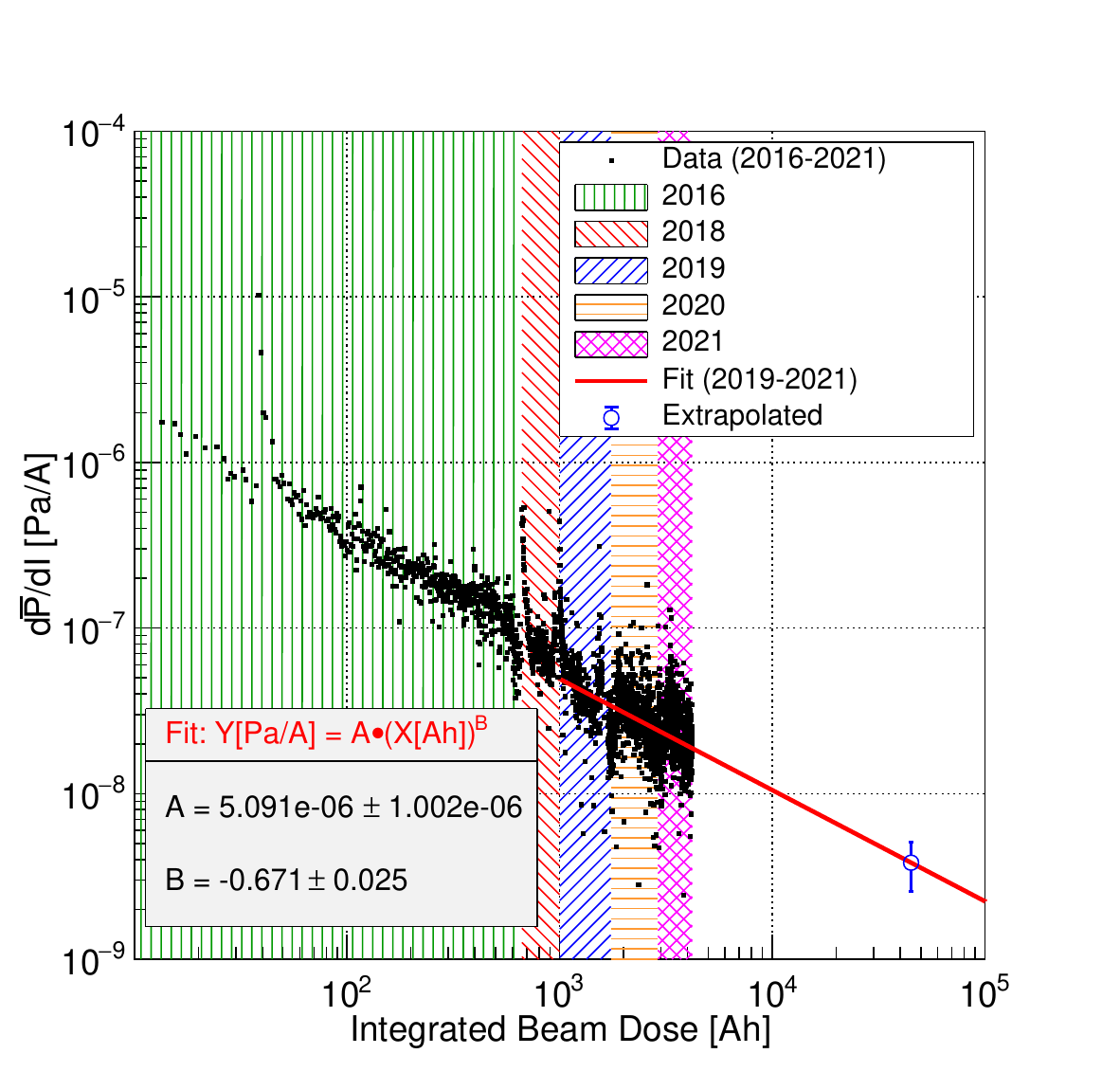}}
\caption{\label{secE:fig1}Beam pipe pressure increase per unit current, $\mathrm{d} \bar{P}/\mathrm{d} I = 3(\bar{P}_\mathrm{CCG} - \bar{P}_\mathrm{0,CCG})$, versus integrated beam dose for (a) the LER and (b) the HER. We assume $\bar{P}_\mathrm{0,CCG} = \SI{10}{nPa}$. Measurements are shown as black squares.}

\end{figure*}

For the extrapolation of the residual gas pressure in each ring, we use the pressure measured by the CCGs to estimate the dynamic pressure evolution. The data were collected throughout the commissioning of SuperKEKB from 2016 until mid-2021. 
Figure~\ref{secE:fig1} shows the estimated average ($\mathrm{d} \bar{P}/\mathrm{d} I$), ring pressure increase per unit current at the center of the beam pipe, versus integrated beam dose ($BD_\mathrm{int.}$).
Each calendar year of operation is emphasized with a different color and hatching style. 
To estimate the dynamic pressure at the beam parameters before LS2 at a luminosity of \SI{2.8e35}{cm^{-2}.s^{-1}}, we fit only the Phase~3 (2019--2021) data. We assume that $BD_\mathrm{int.} = \SI{45}{kAh}$\footnote{Rough estimate made in June~2020, which assumes that the beam is always stored at the maximum beam current ($\sqrt{I_\mathrm{LER}I_\mathrm{HER}}$) during operation, excluding some start-up days in each run period.} will be reached by 2027 at beam currents of \SI{2.52}{A} and \SI{1.82}{A} for the LER and HER, respectively (Table~\ref{secE:tab2}). Blue, open circles in Fig.~\ref{secE:fig1} show the extrapolated pressure increase per unit current for $BD_\mathrm{int.} = \SI{45}{kAh}$. LS2 is currently planned for 2027, but there is significant uncertainty, and it may take longer to reach the integrated beam dose of \SI{45}{kAh} assumed in the pressure extrapolation.

The pressure spikes seen at the beginning of each year are due to compromising the vacuum in short ring sections as part of machine maintenance work performed during standard machine shutdown periods. However, dedicated vacuum scrubbing runs, immediately after each intervention, reduce the pressure down to the nominal level. In the early stages of SuperKEKB commissioning in 2016, beam size blow-up and a non-linear residual gas pressure rise with beam current were observed in the LER~\cite{Ohmi2020, Suetsugu2020}. The growth of the positron beam emittance was caused by a fast head-tail instability, which was induced by the electron cloud effect. In 2018, this effect was cured by attaching permanent magnets and solenoids to most of the beam pipes at drift spaces in the LER. Therefore, a steep change in $\mathrm{d}\bar{P}/\mathrm{d}I$ is seen between 2016 and 2018 in Fig.~\ref{secE:fig1}.

Assuming the base pressure for both rings is at the level of $\bar{P}_\mathrm{0} = \SI{10}{nPa}$, we can calculate the expected value of the beam pipe gas pressure as $\bar{P}_\mathrm{eff.} = \bar{P}_\mathrm{0} + \mathrm{d}\bar{P}/\mathrm{d}I \times I$. Table~\ref{secE:tab1} lists all results of the gas pressure extrapolation, which are then used to normalize the beam-gas background simulation.

\begin{table}[htbp]
\centering
    \caption{\label{secE:tab1}Expected beam pipe gas pressure at the beam parameters before LS2 at $\mathcal{L} = \SI{2.8e35}{cm^{-2}.s^{-1}}$, where $\mathrm{d}\bar{P}/\mathrm{d}I$, $\bar{P}_\mathrm{0}$ and  $\bar{P}_\mathrm{eff.}$ stand for the ring average pressure increase per unit current, base pressure and beam pipe pressure, respectively.}
    \begin{tabular}{ccc}
    \hline\hline
    Term & LER & HER \\
    \hline
    $\mathrm{d}\bar{P}/\mathrm{d}I$~[nPa/A] & $14.94 \pm 4.83$ & $3.83 \pm 1.27$\\
    $\bar{P}_\mathrm{0}$~[nPa] & 10 & 10\\
    $\bar{P}_\mathrm{eff.}$~[nPa] & $47.66 \pm 12.17$ & $16.97 \pm 2.31$\\
    \hline\hline
    \end{tabular}
\end{table}

To simulate the expected beam-gas background at $\mathcal{L}$\,=\,\SI{2.8e35}{cm^{-2}.s^{-1}}, we use the measured gas pressure distribution along each ring from June~2021, shown in Fig.~\ref{secB:fig2}, and scale it to the expected vacuum pressure as follows

\begin{equation}
    P^\mathrm{est.}_\mathrm{CCG, i} = P^\mathrm{meas.}_\mathrm{CCG, i} \times \frac{3\bar{P}_\mathrm{0} + \mathrm{d}\bar{P}/\mathrm{d}I \times I}{3\bar{P}^\mathrm{meas.}_\mathrm{CCG}},
    \label{secE:eq1}
\end{equation}
where $P^\mathrm{est.}_\mathrm{CCG, i}$ and $P^\mathrm{meas.}_\mathrm{CCG, i}$ are the estimated and measured gas pressure at the \textit{i}-th CCG; $\bar{P}_\mathrm{0}$, $\mathrm{d}\bar{P}/\mathrm{d}I$ and $I$ are taken from Table~\ref{secE:tab1}; $\bar{P}^\mathrm{meas.}_\mathrm{CCG}$ is the average ring pressure measured by CCGs; the factor 3 is used to take into account the vacuum conductance between the beam pipe and CCGs, see Section~\ref{subsec:SingleBeam}.

\subsection{Collimation system settings}

For future beam optics and beam parameters, the collimation system must be re-optimized in order to effectively protect the detector from stray beam particles. The optimization procedure~\cite{Nakayama2012,Natochii2021} is based on finding a compromise between very tight collimator apertures, which reduce the beam lifetime and induce beam instabilities, and wide apertures, which increase the beam backgrounds in the IR. One of the instabilities limiting aperture tightening is TMCI, which is a wake-field effect from bunched charges traveling through the machine aperture, causing a strong head-tail instability and beam size increase. We adjust the apertures of all currently installed collimators, see Fig.~\ref{secIntro:fig1}, to satisfy the requirements listed below while maintaining the lowest possible IR backgrounds and beam lifetimes of the order of \SI{15}{minutes} for both rings.

\paragraph{TMCI limits relaxation} 
To satisfy TMCI limits in the LER, we fully open the collimator D03V1 and set D06V2 at the aperture of the IR.
    
\paragraph{Far distant high beam losses}
We perform primary collimation as far as possible from the IR, to reduce secondary showers reaching the detector and to protect the QCS against an abnormally injected beam. Thus we use D06V1 and D02H1 in the LER, and  D09V1/3 and D12H1/2 in the HER.
    
\paragraph{Background reduction around the IR} 
Since tip-scattered particles from the collimators closest to the IP may contribute to the IR background, we shadow these collimators by tightening other upstream collimators, thereby reducing beam losses around the IR. We thus set D02H2 narrower than D02H4 in the LER, and D01H3 narrower than D01H5 in the HER. This configuration should also reduce the neutron flux toward Belle~II from the closest collimators.

The optimized collimators satisfy the TMCI requirement for the predicted bunch currents before LS2 at the luminosity of \SI{2.8e35}{cm^{-2}.s^{-1}}, which are $I^\mathrm{LER}_\mathrm{b} = \SI{1.60}{mA}$ and $I^\mathrm{HER}_\mathrm{b} = \SI{1.15}{mA}$. The maximum allowed bunch currents before reaching instabilities due to collimator and IR beam pipe apertures are $I^\mathrm{LER}_\mathrm{thresh.} = \SI{1.76}{mA}$ and $I^\mathrm{HER}_\mathrm{thresh.} = \SI{1.66}{mA}$ for the LER and HER, respectively. 

\subsection{Predicted Belle~II backgrounds}

\begin{figure*}[htbp]
\centering
\includegraphics[width=\linewidth]{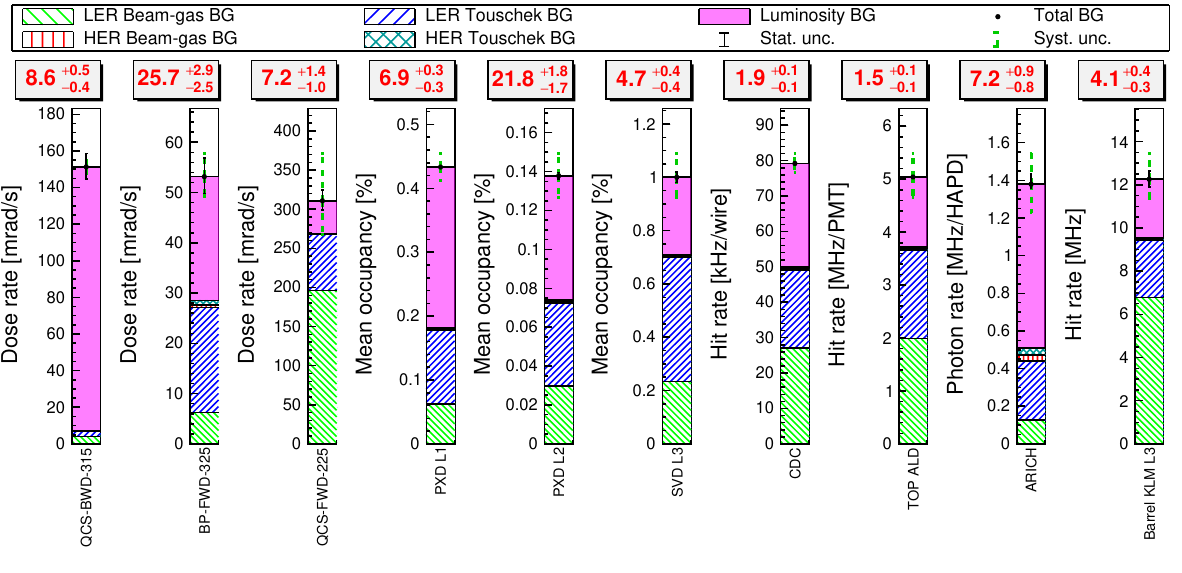}
\caption{\label{secE:fig2}Estimated Belle~II background composition for predicted beam parameters before LS2. Each column is a stacked histogram. The red numbers in rectangles are detector safety factors, showing that Belle~II should be able to operate safely until a luminosity of \SI{2.8e35}{cm^{-2}.s^{-1}}, with some important caveats, discussed in the text.}

\end{figure*}

Figure~\ref{secE:fig2} shows the predicted beam background composition in Belle~II at the beam parameters before LS2 at the luminosity of \SI{2.8e35}{cm^{-2}.s^{-1}}, see Tables~\ref{secE:tab2} and \ref{secE:tab1}. To obtain the expected background rates, each simulated background component is scaled by corresponding Data/MC ratios shown in Fig.~\ref{secD:fig2}. The predictions include systematic uncertainties associated with the variation of the Data/MC ratios among detector layers, sensors or modules. The predicted background is well below the detector limits listed in Table~\ref{secD:tab1}, with safety factors ranging from $\sim 2$ to $\sim 30$, leaving some margin for the injection background and unexpected beam losses. In addition, the usage of the Crab-Waist scheme at $\beta^{*}_\mathrm{y} = \SI{0.6}{mm}$ potentially can enlarge the margin by an additional factor of three, as discussed above.  

\subsection{Predicted neutron flux near Belle~II}

The neutron flux inside Belle~II is currently being studied, and detailed findings will be published separately in the future. Here, we only provide rough estimates, based on older, completed studies in the machine tunnel.

To roughly estimate the neutron fluence in the accelerator tunnel, we assume that single-beam losses at the collimators closest to the IP are well controlled by adjusting upstream collimators. Therefore, we focus on luminosity-production of neutrons only. Based on the TPC data~\cite{Schueler2021}, the 1-MeV neutron equivalent fluence per Snowmass year, at a luminosity of \SI{2.8e35}{cm^{-2}.s^{-1}} is about \SI{5e10}{n_\mathrm{eq}/cm^2} and \SI{5e9}{n_\mathrm{eq}/cm^2} in the BWD and FWD tunnels, respectively~\cite{Schueler2021}. At the same luminosity, our simulation predicts a \ce{^3He}~tube count rate due to thermal neutrons of only about \SI{2e4}{n/(cm^2.s)} and \SI{4e3}{n/(cm^2.s)} in the BWD and FWD tunnels, respectively. The Belle~II limit for the neutron fluence ranges from $10^{12}$ to  $10^{14}~\mathrm{n_{eq}/cm^2}$, as discussed in Section~\ref{sec:CurrentBackgroundLevelsAndMargin}. Hence our current estimates show that the expected neutron background allows safe detector operation for more than 10~years.

\subsection{Planned background mitigation via improved shielding}

During LS1, we plan to install new or improved background shielding. We plan additional neutron shielding of Belle~II to suppress the flux of neutrons originating from the accelerator tunnel and from the QCS. Although below the strict detector limits, the neutrons cause the ageing of ECL photodiodes and other detector components and lead to operationally disruptive SEU events. An additional IR bellows shield made of tungsten, and modified FWD and BWD QCS head plates, currently made of tungsten and planned to be replaced by stainless steel, are under construction and may be installed with the new pixel detector during LS1. This should reduce single-beam and luminosity backgrounds by up to 50\%~\cite{Nakamura2021}. A new IP beam pipe with an additional gold layer and slightly modified geometry to reduce the amount of the back-scattered SR is also in production.

\subsection{Background beyond LS2}

While the Belle~II backgrounds are under control, and their evolution at higher luminosity looks promising, there are other difficulties related to achieving stable machine operation while keeping acceptable background levels. In Ref.~\cite{Natochii2022}, we review ongoing activities and options for further background mitigation, and background predictions for even higher luminosities, up to \SI{6.3e35}{cm^{-2}.s^{-1}}.

\section{\label{sec:Conclusion}Conclusions}

We have reported on the current beam-induced background levels in Belle~II, demonstrated the accuracy of our background predictions, and estimated backgrounds for future SuperKEKB beam parameters. The beam loss simulation software, based on SAD and Geant4, has been significantly improved compared to the versions used in earlier commissioning phases, and now accurately describes the measured detector backgrounds, with Data/MC ratios within one order of magnitude of unity. 

We want to stress that it is crucial to understand all main sources of beam losses affecting machine and detector components' longevity and causing detector performance degradation. Therefore, the accurate background prediction at the current stage is essential to trust any extrapolations, including simulation-based studies of potential SuperKEKB or Belle~II upgrades. We correct the simulation for any remaining discrepancy with measurements by using the Data/MC ratios for re-scaling the simulation. But as opposed to what we had in the past at Phase~1 (2016) and Phase~2 (2018), these correction factors are now much closer to the unity, significantly increasing confidence in our methodology and extrapolations.

In early Phase~3, 
backgrounds from collisions of two beams at the IP, which are expected to dominate at higher luminosities, are slightly ($\sim\!20\%$) lower than expected. Backgrounds from single beams, which currently dominate, are a factor of four different from expectations, which is in line with the size of typical machine systematics involved, such as the beam pipe gas composition, unknown machine errors, beam instabilities, beam-beam effects, and modeling accuracy of machine components and detector surroundings. 

At the current and future stages of the experiment, the most vulnerable sub-detectors are TOP and CDC, whose PMT lifetime and charged tracks reconstruction performances are strongly affected by high beam losses in the IR, respectively. Their safety factors are estimated to be at the level of $\sim 2$ for a luminosity of \SI{2.8e35}{cm^{-2}.s^{-1}}, leaving some margin for unpredicted or imperfectly controlled beam losses.

Currently, the most dangerous backgrounds are due to Touschek and beam-gas scattering in the LER. However, we expect that a further increase of the collision rate above \SI{1e35}{cm^{-2}.s^{-1}} will raise the luminosity background to the same level as single-beam backgrounds. Based on our measurements and current understanding of beam loss mechanisms in SuperKEKB, we predict that as beam currents are increased and the beam size is decreased in the next decade, beam-induced backgrounds in Belle~II will remain acceptable until at least $\mathcal{L} = \SI{2.8e35}{cm^{-2}.s^{-1}}$ at $\beta^{*}_{\rm y}=\SI{0.6}{mm}$. This statement assumes the baseline plan of replacing the short-lifetime conventional and ALD MCP-PMTs in the TOP detector, stable and well-controlled main ring and injection chain operation, continuous progress on vacuum scrubbing, and low impact from beam instabilities. Installing additional shielding during the two long shutdowns in 2022--2023 and around 2027 could reduce backgrounds further. 

There are several important uncertainties in our projections of future backgrounds, such as unexpected and uncontrolled catastrophic beam losses, unknown sources of machine impedance, vacuum pressure at high beam doses, and possible IR beam pipe upgrades. These issues could affect our background forecast in either direction and require further studies and refinement.

Backgrounds from neutrons have been studied with dedicated detectors in the SuperKEKB tunnel. While the flux appears understood and manageable in the short term, a quantitative study that connects neutron rates to Belle~II hit and SEU rates is needed, ongoing, and will be published separately in the future. SEUs deserve special scrutiny as they can reduce the operational efficiency of the experiment.

Backgrounds from injection also appear manageable but have not been projected forward, as they are not simulated from first principles. This is a challenging task that should also be tackled in the future. Machine learning techniques appear useful in identifying the injection background, could be helpful in online machine diagnostics and may detect the most crucial parameters to be adjusted for background mitigation and collider performance improvement.

Mainly due to the uncertainties related to the design machine lattice and beam instabilities, it is too early to make accurate predictions for the distant future, but backgrounds could exceed detector limits at $\mathcal{L} = \SI{6e35}{cm^{-2}.s^{-1}}$ for $\beta^{*}_{\rm y}=\SI{0.3}{mm}$. Thus, several machine operation schemes, instability and background countermeasures, and upgrades of the experiment are under consideration in order to collect an integrated luminosity of the order of \SI{50}{ab^{-1}} by the 2030s. We are closely collaborating with EU, US and Asian accelerator laboratories on optimizing upgrades of SuperKEKB and reaching the target luminosity.

\section*{Acknowledgements}

We thank the SuperKEKB accelerator and optics groups for the excellent machine operation and for sharing their lattice files; the KEK cryogenics group for the efficient operation of the solenoid; the KEK computing team for on-site support; our Belle~II colleagues for the detector operation; G.~Casarosa (INFN~Pisa), D.~E.~Jaffe (BNL), K.~Oide (CERN), B.~Spruck (JGU~Mainz), K.~Trabelsi (IJCLab), Y.~Funakoshi, N.~Iida, T.~Koga, H.~Koiso, G.~Mitsuka, K.~R.~Nakamura, Y.~Ohnishi, Y.~Suetsugu and D.~Zhou (KEK) for their ideas, assistance, comments on the paper and constructive discussions. This work was supported by the U.S. Department of Energy (DOE) via Award Numbers DE-SC0010504, DE-SC0010007, and DE-SC0019230 and via U.S. Belle~II Operations administered by Brookhaven National Laboratory (DE-SC0012704); the National Institute of Informatics, and Science Information NETwork~5 (SINET5), and the Ministry of Education, Culture, Sports, Science, and Technology (MEXT) of Japan. This project has received funding from the European Union’s Horizon 2020 Research and Innovation programme under Grant Agreement (GA) No.~654168 and GA No.~101004761. We acknowledge the support of Grant CIDEGENT/2018/020 of Generalitat Valenciana (Spain).

\bibliography{mybibfile}

\end{document}